\documentclass[conference]{IEEEtran}

\usepackage{color,array,amsthm}
\usepackage{graphicx}
\usepackage{cite}
\usepackage{amsmath,amssymb,amsfonts}
\usepackage{algorithm}
\usepackage{algpseudocode}
\usepackage{graphicx}
\usepackage{textcomp}
\usepackage{xcolor}
\graphicspath{ {images/} }
\usepackage{makecell} 

\begin{document}

		\title{Contact Plan Design for Cross-Linked GNSSs:\\An ILP Approach for Extended Applications}
	
	\author{
		\IEEEauthorblockN{
			Huan Yan\IEEEauthorrefmark{1},
			Juan A. Fraire\IEEEauthorrefmark{2}\IEEEauthorrefmark{3}
			Ziqi Yang\IEEEauthorrefmark{1},
			Kanglian Zhao\IEEEauthorrefmark{1},
			Wenfeng Li\IEEEauthorrefmark{1},\\
			Xiyun Hou\IEEEauthorrefmark{4},
			Haohan Li\IEEEauthorrefmark{4},
			Yuxuan Miao\IEEEauthorrefmark{4},\\
			Jinjun Zheng\IEEEauthorrefmark{5},
			Chengbin Kang\IEEEauthorrefmark{5},
			Huichao Zhou\IEEEauthorrefmark{5},
			Xinuo Chang\IEEEauthorrefmark{5},
			Lu Wang\IEEEauthorrefmark{5},
		}
		
		\IEEEauthorblockA{
			\IEEEauthorrefmark{1}Nanjing University, Nanjing, China\\
			\IEEEauthorrefmark{2}Inria, INSA Lyon, CITI, UR3720, 69621 Villeurbanne, France\\
			\IEEEauthorrefmark{3}CONICET -- Universidad Nacional de Córdoba, Córdoba, Argentina\\
			\IEEEauthorrefmark{4}Nanjing University Shcool of Astronomy and Space Science, Nanjing, China\\
			\IEEEauthorrefmark{5}China Academy of Space Technology (CAST)
	}}
	
	\maketitle

	\begin{abstract}Global Navigation Satellite Systems (GNSSs) employ inter-satellite links (ISLs) to reduce dependency on ground stations (GSs), enabling precise ranging and communication across satellites. 
	Beyond their traditional role, ISLs can support extended applications, including providing navigation and communication services to external entities. 
	However, designing effective contact plan design (CPD) schemes for these multifaceted ISLs, operating under a polling time-division duplex framework, remains a critical challenge. 
	Existing CPD approaches focus solely on meeting GNSS satellites' internal ranging and communication demands, neglecting their extended applications.
	This paper introduces a CPD scheme capable of supporting extended GNSS ISLs. 
	By modeling GNSS requirements and designing a tailored service process, our approach ensures the allocation of essential resources for internal operations while accommodating external user demands. 
	Based on the BeiDou constellation, simulation results demonstrate the proposed scheme's efficacy in maintaining core GNSS functionality while providing extended ISLs on a best-effort basis. 
	Additionally, the results highlight the significant impact of GNSS ISLs in enhancing orbit determination (OD) and clock synchronization for the Earth-Moon libration point (LP) constellation, underscoring the importance of extended GNSS ISL applications.
	\end{abstract}
	
	
	\section{INTRODUCTION}
	\label{sec_introduction}
	
	Global Navigation Satellite Systems (GNSSs) are satellite-based infrastructures that provide global Positioning, Navigation, and Timing (PNT) services~\cite{50}. 
	Prominent operational constellations include the United States GPS, Russia’s GLONASS, Europe’s Galileo, and China’s BeiDou Navigation Satellite System.
	
	To reduce dependency on ground stations (GSs), GNSS architectures have integrated Inter-Satellite Links (ISLs)~\cite{51,52,53}, allowing satellites outside GS line-of-sight to be monitored effectively. 
	ISLs enable precise ranging and data exchange across the constellation, supporting accurate orbit determination (OD) and time synchronization~\cite{9}. 
	Additionally, ISLs expand the functional capabilities of GNSS, paving the way for diverse extended applications.
	The extended applications enabled by GNSS ISLs include the following scenarios:
	\begin{enumerate}
		\item \textbf{Integration with other PNT infrastructures}: GNSS establishes ISLs with systems such as Earth-Moon libration points (LPs) constellations~\cite{10} and low Earth orbit GNSS augmentation constellations~\cite{11,13}. This integration aims to create a cohesive and precise PNT infrastructure that leverages multiple sources~\cite{2,4}.
		\item \textbf{Enhanced navigation and timing services}: 
		In the context of Lunar navigation for example, ISLs between GNSS and spacecraft provide significantly more precise navigation and timing services~\cite{23,24} with respect to GNSS sidelobe signals~\cite{54}.
		\item \textbf{Data transmission and communication relay}: Leveraging its data transmission capabilities~\cite{9}, GNSS ISLs can act as communication access points for external spacecraft, facilitating data relay to GSs~\cite{36}.
	\end{enumerate}
	
	In light of ISLs' benefits to GNSS and its extended applications, a practical approach to constructing ISLs involves using RF phased-array narrow beams within a polling time-division duplex system~\cite{27,32}. 
	Under this framework, satellites establish ISLs on a time-slot basis. 
	
	To achieve high-precision OD and time synchronization, each satellite within the GNSS network must establish ISLs with a sufficient number of distinct neighbors, ensuring access to diverse-ranging data. 
	Due to platform constraints, each satellite is typically equipped with a single narrow-beam terminal, allowing it to establish an ISL with only one object at a time~\cite{32,45,9}. 
	This limitation implies that complete end-to-end communication paths can not always be formed within a single time slot. 
	Instead, such paths are created over multiple successive time slots to facilitate data transmission.
	
	Additionally, modern and future GNSS aims to extend its ISL capabilities to support \textit{extended applications} in external entities such as satellites at the Earth-Moon LPs, spacecraft operating in medium and high Earth orbits, and low Earth orbit augmentation satellites for the GNSS system. 
	All these external entities are collectively referred to in this paper as \textit{users}. 
	These users demand ISLs based on their unique needs, often conflicting with GNSS’s core positioning requirements, creating a challenging scheduling problem for ISL allocation.
	
	Designing the dynamic GNSS topology over time slots to satisfy ranging requirements, enable end-to-end data flows, and support extended applications is the core challenge addressed by Contact Plan Design (CPD)~\cite{37}.
	In essence, CPD involves selecting an optimal subset of links, referred to as contacts, from a pool of potential links. 
	This selection is made with the objective of maximizing system performance (e.g., system throughput) while satisfying specified constraints (e.g., visibility requirements). 
	
	Previous GNSS CPD solutions have primarily focused on achieving diversified ranges and minimizing communication latency~\cite{17,31,32,27,44,45,30,46,47,48,49}. 
	However, these approaches have overlooked the necessity of supporting the extended applications of GNSS ISLs. 
	Implementing these extended applications introduces several new challenges:
	\begin{enumerate}
		\item Limited time-slot resources available for extending ISL applications while ensuring the resources required for GNSS ranging and communication.
		\item The time-division nature of ISLs complicates providing continuous services to users.
		\item Addressing diverse user requirements within the existing GNSS operational framework constraints.
	\end{enumerate}
	
	Building on the current research landscape and these challenges, we assert that developing CPD schemes that simultaneously fulfill GNSS’s internal ranging and communication needs while efficiently supporting extended applications is essential for advancing next-generation GNSS systems. 
	To our knowledge, this work is the first to propose and tackle this issue.
	The contributions of this paper are summarized as follows:
	\begin{enumerate}
		\item \textbf{Service Procedure for GNSS User Needs:} We model user requirements and, combined with the GNSS topology model, propose a service procedure to address diverse user needs during GNSS operations. Importantly, we highlight GNSS as infrastructure for external users, who can choose to accept or decline the ISLs offered by GNSS. User operations remain isolated from GNSS operations apart from the established ISLs.
		\item \textbf{CPD Schemes for Extended GNSS ISLs:} This paper explores CPD schemes that extend the applications of GNSS ISLs while meeting internal ranging and communication requirements. We formulate the problem as an ILP (Integer Linear Programming) model, prioritizing user demands while maximizing GNSS system throughput. This approach facilitates continuous ISLs for users across multiple consecutive time slots, addressing the challenge of providing uninterrupted service to a significant extent.
		\item \textbf{Performance Evaluation of Proposed Algorithm:} Extensive simulations were conducted to evaluate the performance of the proposed algorithm. First, we identified the optimal ILP parameter configurations for the discussed scenarios. Next, we assessed the extended capacity of the GNSS system and simulated its performance with extended ISLs over a seven-day period. Finally, we analyzed the impact of GNSS-provided ISLs on the Earth-Moon LP constellation. The results show that our algorithm efficiently meets expansion requirements while improving GNSS performance.
	\end{enumerate}
	
	The remainder of this paper is organized as follows:  
	Section~\ref{sec_background} reviews related work, summarizing CPD methodologies and the extended applications of GNSS ISLs.  
	Section~\ref{sec_system_model} details the GNSS topology model and the user concept employed in this study. It also describes the service process developed within the GNSS framework to address diverse user requirements.  
	Section~\ref{sec_ilp_based_gnss_cpd} introduces an ILP-based GNSS CPD model. This approach is designed to provide users with continuous ISLs spanning single or multiple time slots while ensuring the availability of GNSS ranging and communication resources.  
	Section~\ref{sec_evaluation} presents a performance evaluation of the proposed ILP-based GNSS CPD model, including simulations assessing the impact of GNSS ISLs at different frequencies on OD and clock synchronization for the Earth-Moon LP constellation.  
	Finally, Section~\ref{sec_conclusion} concludes the paper with a summary of findings.
	Furthermore, to assist readers in understanding the paper, we have compiled a list of acronyms in Table~\ref{acronyms}.
		\begin{table}[]
		\centering
		\caption{List of Acronyms}
		\label{acronyms}
		\begin{tabular}{|l|c|}
			\hline
			\textbf{Acronym} & \textbf{Definition} \\
			\hline
			GNSSs  & Global Navigation Satellite Systems \\
			\hline
			ISLs & Inter-Satellite Links \\
			\hline
			CPD  & Contact Plan Design \\
			\hline
			GSs & Ground Stations \\
			\hline
			LPs & Libration Points \\
			\hline
			OD & Orbit Determination \\
			\hline
			ILP & Integer Linear Programming\\
			\hline
			PNT & Positioning, Navigation, and Timing\\
			\hline
			DRO &Distant Retrograde Orbit \\
			\hline
			EIRP & Equivalent Isotropically Radiated Power\\
			\hline
		\end{tabular}
	\end{table}
	\section{Background}
	\label{sec_background}
	
	This section reviews the context and existing scholarly work on the extended applications of GNSS ISLs and the corresponding CPD techniques.
	
	\subsection{Extended Applications}
	
	\subsubsection{Feasibility}
	The feasibility of extended applications for GNSS ISLs hinges on the system's inherent capacity and payload design.
	
	\paragraph{Expansion Capacity}
	The study in~\cite{8} examines the expandability of the BeiDou-3 system across multiple dimensions, including platform carrying capacity, equipment installation, layout optimization, power margins, thermal control and heat dissipation surfaces, and remote control data transmission interfaces. 
	These factors collectively establish a solid foundation for broadening GNSS functionalities, enabling the system to accommodate extended applications.
	
	\paragraph{Time Slot Resource Margin}  
	The study in~\cite{17} compares ISL scheduling statistics across three CPD methods. 
	It highlights that even while fulfilling the ranging and communication needs of BeiDou satellites, a portion of ISLs terminals remain idle.
	These idle time slots present an opportunity to support external applications.
	
	\paragraph{Earth-Moon Link Budget}  
	Establishing ISLs necessitates meeting two key conditions: (1) geometric visibility and (2) received signal strength exceeding the receiver sensitivity~\cite{55}.  
	To illustrate GNSS ISLs' adequacy in link budget for supporting the most demanding extended applications, we consider their use in Cislunar space navigation.
	Cislunar space refers to the region beyond Earth's atmosphere that is influenced by the Earth–Moon system's gravitational field. It can be approximated as a spherical volume centered on Earth, with a radius of approximately 450,000 km, encompassing the Moon~\cite{66}.  
	Specifically, we focus on the propagation link between GNSS satellites and lunar orbiters, showcasing their capability to meet the stringent requirements of such scenarios.
	
	The quality of the received signal is typically characterized by the Carrier-to-Noise Density Ratio ($C/N_0$). 
	This key metric depends on the received power and the noise level in the reception environment, independent of the receiver's noise bandwidth. 
	The calculation is expressed as:
	\begin{equation}
		C/N_0 = P_r - 10\log_{10}(T_{sys}) + 228.6 - L_{ADC}, \label{eq1}
	\end{equation}
	where $T_{sys}$ represents the equivalent system noise temperature, set to 290~K; 228.6 is the Boltzmann constant in dB, which has the unit of dBW/Hz; $L_{ADC}$ denotes the signal quantization loss after A/D conversion, assumed to be 3~dB; and $P_r$ is the received signal power. 
	The received signal power is defined as:
	\begin{equation}
		P_r = EIRP + L_f + G_r, \label{eq1_added}
	\end{equation}
	where $EIRP$ represents the equivalent isotropically radiated power of the transmitting satellite in the main lobe direction, $G_r$ is the gain of the receiving antenna in the main lobe direction, and $L_f = 20\log_{10}(\frac{\lambda}{4\pi d})$ is the free-space propagation loss, where $\lambda$ is the signal wavelength and $d$ is the propagation distance. 
	Given the high orbital altitudes of GNSS and lunar-orbiting satellites, atmospheric losses are considered negligible and excluded from the calculations.
	
	\begin{table}[]
		\centering
		\caption{Earth-Moon Link Budget}
		\label{link_budget}
		\begin{tabular}{|l|c|}
			\hline
			\textbf{Parameter} & \textbf{Value} \\
			\hline
			Frequency (GHz) & Ka-band (26.5 $\sim$ 40) \\
			\hline
			EIRP (dBW) & 46 $\sim$ 48 \\
			\hline
			$L_f$ (dB) & 233.97 $\sim$ 237.54 \\
			\hline
			$G_r$ (dBi) & 25 $\sim$ 27 \\
			\hline
			$P_r$ (dBW) & -165.95 $\sim$ -161.37 \\
			\hline
			$C/N_0$ (dB-Hz) & 34.44 $\sim$ 42.01 \\
			\hline
		\end{tabular}
	\end{table}
    A gain of 30 dB is common for spaceborne Ka-band phased array antennas~\cite{63,64,65}, while the Ku-Ka band phased array antenna onboard Starlink achieves an even higher gain of 37.1 dB~\cite{62}.
    Based on these observations, we conservatively estimate the gain of the ISL phased array antenna onboard GNSS satellites to be 25–27 dBi.
    The mainlobe gain of current GNSS Earth-directed antennas is approximately 14 dBi~\cite{61,20}, and the EIRP of BDS-III is 35 dBW~\cite{60}. 
    Assuming that the transmission power for the GNSS ISL is the same as that for the Earth-directed beam, the EIRP for the GNSS ISL can be inferred as 46–48 dBW. 
    
    Since bidirectional measurements are required between users and GNSS satellites via the ISL, the user terminal payload should be roughly symmetric with the GNSS satellite payload. Therefore, the receiving antenna gain $G_r$ is also taken as 25–27 dBi.
    
    As the scope of users considered in this paper is limited to the cislunar space, a maximum propagation distance of 450,000 km-the radius of the cislunar space sphere-is adopted in this analysis. At this distance, the path loss in the Ka band ranges from 233.97 dB to 237.54 dB. If the link budget meets the required specifications at this distance, then the link budget between GNSS satellites and users located anywhere within the cislunar space will be sufficient.
    
    Integrating all data, the calculated link budget is presented in Table~\ref{link_budget}.
	The minimum value of $C/N_0$ is 34.44 dB-Hz, which exceeds the signal tracking capabilities of the NASA-developed Navigator receiver. 
	The current generation of the Navigator receiver can maintain stable tracking for signals at levels as low as 22-25 dB-Hz~\cite{34}. 
	In comparison, the next-generation version is anticipated to lower this threshold to 10-12 dB-Hz~\cite{35}. 
	Hence, the link is deemed viable for the intended extended application in the Earth-Moon context.

	\subsubsection{Applicability}
	The extended application scenarios of GNSS ISLs, reflecting the diverse roles GNSS can play in providing ISLs to external users, can be broadly categorized into the following three classes:
	
	\paragraph{Comprehensive PNT}
	GNSS spatiotemporal information services face inherent vulnerabilities due to weak signal strength, limited penetration, and high susceptibility to interference. 
	The concept of \textit{comprehensive PNT} has been proposed to mitigate these challenges~\cite{2,4}. 
	Comprehensive PNT expands the scope of PNT information sources by leveraging diverse physical principles to develop a resilient and complementary multi-source information system. 
	Sources include navigation constellations at LPs, GNSS, LEO navigation augmentation constellations, ground-based navigation enhancement station networks, indoor positioning beacon networks, surface maritime positioning buoy networks, and undersea acoustic positioning beacon networks, among others~\cite{3}.
	
	Numerous studies have explored the interaction between GNSS and other PNT infrastructures through ISLs within comprehensive PNT systems. 
	The findings in~\cite{10} reveal that the Earth-Moon LP constellation, by establishing ISL ranging with the BeiDou satellites, can ensure traceability and alignment of the Earth-Moon spatiotemporal reference with the high-precision Earth-based spatiotemporal reference. 
	Similarly,\cite{19} highlights that in the LP  and BeiDou satellites scenario, augmented ISLs with the BeiDou system significantly improve the orbital accuracy of the LP constellation and reduce the OD arc length. 
	The work in~\cite{13,11} demonstrates that integrating OD for LEO satellites with that of BeiDou MEO and GEO satellites enhances OD accuracy for BeiDou satellites. 
	This enhancement indirectly strengthens the PNT service capabilities of the BeiDou System.
	Moreover, the microgravity environment of a space station fosters superior atomic clock performance, which, as reported in~\cite{58}, allows the station's time reference to be transferred via ISL to BeiDou satellites for dissemination to a wider user community.
	
	\paragraph{Cislunar Spacecraft Navigation}
	The study in~\cite{24} introduces a fast positioning algorithm for navigation receivers assisted by GNSS ISLs, enabling rapid positioning of high-orbit spacecraft during their transfer phases in the cislunar orbit. 
	To address the urgent demand for high-precision tracking and reliable cataloging of non-cooperative targets in cislunar space,\cite{23} proposes a measurement method based on GNSS ISLs and Connected Element Interferometry for high-value cislunar space targets. 
	Additionally,\cite{14} explores an autonomous joint OD method that integrates the Queqiao relay satellite from the Chang’e-4 mission with the BeiDou Constellation, offering a novel approach to expanding ISL applications within the BeiDou system.
	
	\paragraph{User Data Transport}
	The study in~\cite{9} highlights that BeiDou-3 can continuously perform precise ISL ranging across its entire constellation while enabling data transmission via ISLs. 
	This integration of communication and navigation link capabilities positions BeiDou as a potential provider of communication services to external space users through its ISL. 
	Furthermore,~\cite{36} demonstrates using BeiDou ISLs for monitoring services during atmospheric re-entry. 
	This capability allows users to transmit telemetry data to GSs via BeiDou satellite relays without compromising the performance of the BeiDou constellation.
	
	Efficiently scheduling GNSS ISLs to balance diverse and competing demands is a critical challenge, necessitating advanced contact planning strategies to optimize link allocation while preserving GNSS performance.
	
	\subsection{Contact Plan Design}
	
	A \textit{contact} represents a temporal opportunity to establish a communication link between two nodes, provided physical prerequisites such as antenna alignment and sufficient received signal power are met. 
	The collection of all feasible contacts within a network over a given time interval forms the \textit{contact topology}. 
	From this topology, a \textit{contact plan} is generated by selecting a subset of contacts defining the communication links to be established. 
	This selection process accounts for various constraints, including interference, power availability, and resource limitations. 
	The optimization of this subset is the essence of Contact Plan Design (CPD)~\cite{37}.

	Before this study, the primary optimization objectives for GNSS CPD were to enhance OD accuracy by acquiring diverse ranging data and to minimize communication delays for expedited telemetry data transmission to GSs.
	The advancements in CPD for navigation have evolved over the years as follows:
	\begin{itemize} 
		\item In 2015, \cite{47} formulated CPD as a constrained optimization problem and solved it using a simulated annealing-based heuristic algorithm.
		\item In 2017, \cite{32} proposed a grouping-based approach to maximize range observations and minimize communication delays between satellites and ground facilities.
		\item In 2018, \cite{27} introduced a three-step link scheduling method that uses a genetic algorithm to identify optimal downlink routes.
		\item Also in 2018, \cite{30} improved inter-satellite ranging and communication performance by balancing bipartite and general matching approaches.
		\item In 2019, \cite{45} introduced the first distributed CPD scheme for GNSS using an algorithmic approach running on each GNSS satellite.
		\item In 2020, our prior work~\cite{17} was the first to formulate CPD in GNSS as an ILP problem, optimizing communication performance under ranging constraints. 
		\item In 2021, \cite{48} employed a non-dominated sorting genetic algorithm to generate and optimize satellite sequences for all time slots, applying the Blossom algorithm to achieve maximum matching.
		\item Also in 2021, \cite{57} proposed a CPD scheme for Optical Inter-Satellite Links based on a degree-constrained minimum spanning tree heuristic. 
		\item In 2022, \cite{31} extended the polling mechanism to include ground-satellite links, minimizing the average data delivery delay from satellites to GSs.
		\item Also in 2022, \cite{46} classified links into four types, adjusting their corresponding weights to optimize performance.
		\item In 2023, \cite{44} developed an adaptive topology optimization algorithm based on signed variance, leveraging the feedback effect of allocated links on subsequent assignments as prior knowledge.
	\end{itemize}
	\subsection{The Gap in CPD for Extended Applications}
	
	Most existing CPD studies focus on internal GNSS objectives, such as improving OD accuracy and reducing telemetry delays, while neglecting the broader demands of external users and extended applications. 
	Despite the demonstrated feasibility and transformative potential of GNSS ISLs for these applications, a critical gap remains in developing strategies that simultaneously support extended applications and preserve GNSS’s core ranging and communication functions.
	This paper addresses this gap by proposing advanced CPD schemes that balance these competing priorities. Specifically, we design adaptive and efficient link allocation strategies to accommodate diverse extended applications while ensuring uninterrupted GNSS performance. Our approach paves the way for a more versatile and future-ready GNSS framework capable of adapting to evolving application demands.

	\section{System Model}
	This section presents the model setup employed in this work, establishing the foundation for the subsequent GNSS CPD.
	\label{sec_system_model}
	
	\subsection{Topology Model}
	
	In satellite networks, two nodes are considered visible to each other if they are unobstructed by celestial bodies and fall within each other’s antenna-pointing angles. 
	This visibility relationship is inherently dynamic due to the continuous motion of satellites. 
	The satellite network is modeled as a Finite State Automaton (FSA)~\cite{18} to address this challenge. 
	The FSA divides the entire scheduling period into a series of equal-length time segments, referred to as \textit{FSA states}, to manage the dynamic nature of node visibility. 
	Within each state, two nodes are defined as visible if they maintain visibility throughout the state duration; otherwise, they are considered invisible. 
	Topology planning is performed based on this static visibility within each state.
	
	Each FSA state is further subdivided into equal-length \textit{superframes}, which serve as the fundamental units for link allocation. 
	To enhance OD accuracy in GNSS, satellites must frequently switch ISLs over short intervals to gather diverse ranging data. 
	GNSS employs a Polling Time Division Duplex scheme to achieve this, dividing each superframe into multiple equal-length time \textit{slots}, typically spanning 3 seconds~\cite{30,27}. 
	These time slots act as the essential time units for establishing ISLs. 
	GNSS achieves greater range diversity and faster data transmission by dynamically establishing ISLs with different satellites in successive time slots.
	
	In addition to ISLs between GNSS satellites, the extended applications referred to as \textit{users} necessitate establishing ISLs with external entities. 
	While user requirements may vary, ISL services to external users are provided in integer multiples of time slots to ensure compatibility with GNSS operations.
	
\begin{figure}[] 
		\centering
		\includegraphics[width=0.8\linewidth]{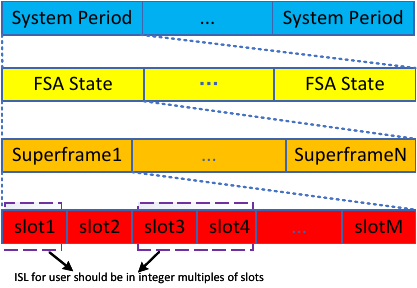}
		\caption{Hierarchical Topology Model: System periods are divided into Finite State Automaton (FSA) states, further segmented into superframes subdivided into time slots as the basic ISL allocation unit.}
		\label{fig1}
	\end{figure}

	Under these definitions, the hierarchical topology model that serves as the foundation for the remainder of this paper is illustrated in Fig.~\ref{fig1}.
	
	\subsection{User Model}
	
	A user may represent an individual spacecraft, such as the Queqiao relay satellite highlighted in~\cite{14}, or an entire constellation, such as the Earth-Moon LP constellation discussed in~\cite{10,19}. 
	Users require tailored ISL services to meet their specific operational needs, including navigation, communication, or other extended applications enabled by GNSS.
	
	\subsubsection{User Preprocessing}
	
	Some {constellation users} require ISLs from GNSS to any node within the constellation without needing links to specific nodes. 
	
    For instance, the Earth–Moon LP constellation, owing to its unique dynamical characteristics, can achieve autonomous OD solely through intra-constellation ranging, assuming negligible clock offsets among the constituent satellites. 
    However, without timing references from GNSS satellites or GSs, the constellation cannot realize absolute time calculation~\cite{67,68}. 
    We then consider leveraging GNSS satellites to provide time-transfer services to the LP constellation. In such a scenario, establishing a link between any GNSS satellite and an arbitrary satellite within the LP constellation enables the dissemination of the GNSS time reference to that LP satellite, which subsequently propagates the time reference throughout the entire LP constellation via internal inter-satellite links. 
    Under this time-transfer architecture, establishing a link between a GNSS satellite and any single LP satellite is sufficient to enable time dissemination.

\begin{figure}[] 
		\centering
		\includegraphics[width=0.85\linewidth]{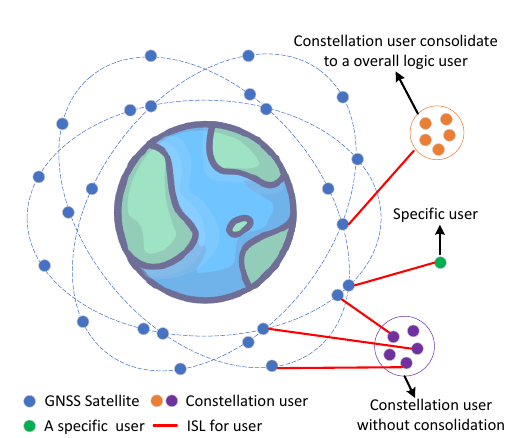}
		\caption{User Model: The diagram illustrates the interaction between GNSS satellites and user types. \textit{Constellation users} suitable for consolidation are represented as a single logical user.}
		\label{fig2}
	\end{figure}

	Given that the complexity of CPD increases with the number of nodes and edges in the topology, constellation users such as the aforementioned LP constellations, which require timing services and are indifferent to which specific node within the constellation establishes a link with a GNSS satellite-can be aggregated into a overall logical user before GNSS CPD process, as illustrated in Fig.~\ref{fig2}. The visibility of this logical user to GNSS is determined by the union of the visibility of all individual nodes within the constellation.
	
	In contrast to the overall logic constellation users, specific users or constellation users requiring node-level differentiation for ISLs illustrated in Fig.~\ref{fig2}, cannot be logically aggregated.

	It should be noted that, as an infrastructure provider, GNSS cannot unilaterally decide whether a constellation user should be treated as a single logical entity. The decision to aggregate a constellation into a logical user must be driven by the user’s own operational requirements and explicitly communicated to the GNSS system. 
	For instance, if a LP constellation requires only timing services, it should inform the GNSS system-when requesting link services-that it is indifferent to which specific LP node establishes the link. In this case, the GNSS system may treat the entire constellation as a single logical user. 
	Conversely, if the LP constellation requests OD services for one or more specific nodes within the constellation, the GNSS system must refrain from logical aggregation and instead maintain strict node-level distinction among the constituent nodes when establishing links and delivering services.

	\subsubsection{User Isolation}
\begin{figure}[] 
		\centering
		\includegraphics[width=0.9\linewidth]{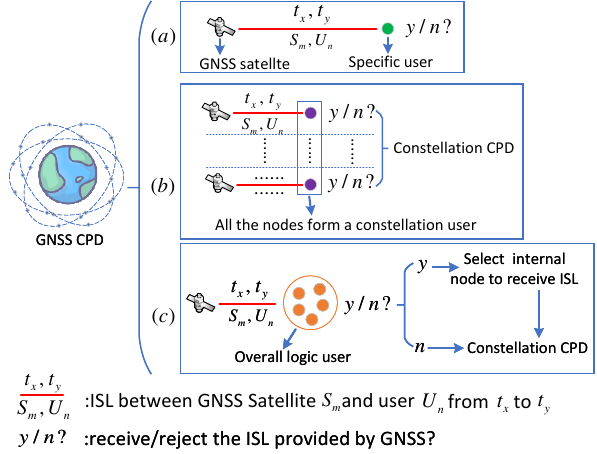}
		\caption{User Isolation: (a) Interaction between a GNSS satellite and a specific user. (b) Constellation users without consolidation, with GNSS providing ISLs to individual nodes and the constellation CPD managing internal links. (c) Consolidated overall logic user, with GNSS offering ISLs to the constellation as a whole, requiring the selection of an internal node to receive ISLs before performing the constellation CPD.}
		\label{fig3}
	\end{figure}

	In this work, users and the GNSS system are treated as distinct operational entities.
	While GNSS establishes ISLs with users, their operations remain independent, as evidenced by the following key aspects:
	
	\begin{enumerate}
		\item \textbf{User Autonomy in ISL Establishment}: GNSS acts as an infrastructure providing ISLs, with its CPD determining the timeslots available for user connections. However, the decision to utilize these timeslots lies entirely with the users, as illustrated by the \textit{y/n?} in Fig.~\ref{fig3}(a). GNSS does not impose ISL establishment during the allocated timeslots.
		\item \textbf{Independence for Unconsolidated Constellation Users}:  For constellation users that are not consolidated (Fig.~\ref{fig3}(b)), the process operates as follows:  
		1) GNSS CPD determines the timeslots for ISLs offered to constellation nodes.  
		2) Based on their strategies, Constellation users decide whether to accept or reject the offered ISLs. 
		3) After deciding, constellation users perform their own CPD to establish internal links, ensuring no conflict with the accepted ISLs from GNSS.
		\item \textbf{Internal Autonomy for Consolidated Logical Users}:
		For consolidated constellation users represented as a single logical entity (Fig.~\ref{fig3}(c)), GNSS CPD provides ISLs for the constellation as a whole. The constellation must then select an appropriate internal node to receive these ISLs, adhering to two conditions:  
		1) The selected node must be visible to the GNSS satellite offering the ISL.  
		2) Accepting the ISL must not disrupt the constellation's internal operations.  
		Once the node is selected, the constellation performs its CPD to configure internal links accordingly.
		\item \textbf{Refinement of GNSS CPD Results Based on User Decisions}:
	    After users decide whether to accept or reject the ISLs offered by GNSS, they must convey this decision to the GNSS system. For rejected links, the GNSS system cancels the corresponding link establishment plan to conserve resources, such as energy. 
		Similarly, when a constellation overall logical user selects specific internal nodes to receive GNSS-provided ISLs, it must inform the GNSS system of these choices to unambiguously identify the GNSS provided-ISL endpoints.
		Once all such user feedback is collected, the GNSS CPD-derived link scheduling results become fully determined.

	\end{enumerate}
	
It should be noted that the aforementioned GNSS CPD process, user decision-making, and feedback processes are assumed to be performed offline on the ground. After these procedures, the resulting contact plan is uplinked by ground stations to both the GNSS constellation and the users, who strictly adhere to this schedule when establishing ISLs.
	
	\subsubsection{User Requirements}
	
	The GNSS CPD provides ISLs to users based on requirements predefined by ground operators. 
	These requirements are formally represented as \scalebox{0.9}{$\vec{U} = {(U_1, [a_1, b_1, c_1, d_1]), \dots, (U_n, [a_n, b_n, c_n, d_n])}$}, where:
	\begin{enumerate}
		\item $U_n$ is the unique identifier for the user.
		\item $[a_n, b_n, c_n, d_n]$ specifies the user’s ISL request:
		\begin{itemize}
			\item $a_n$: occurrence interval (in FSA states) at which the user expects ISLs.
			\item $b_n$: Duration of each ISL (in time slots).
			\item $c_n$: Total number of ISLs requested.
			\item $d_n$: The number of user-carried terminals determines the maximum number of simultaneous ISLs that can be established with GNSS.
		\end{itemize}
	\end{enumerate}
	
	For example, the requirement $(U_1, [3, 2, 4, 1])$ specifies that user $U_1$ expects four ISLs of two time slots each, occurring at intervals of three FSA states (e.g., during the first, fourth, and seventh FSA states). Since $d_1 = 1$, $U_1$ can only establish one ISL with a GNSS satellite at any given time.
	
	\subsubsection{User Service Procedures}
	
\begin{figure}[] 
		\centering
		\includegraphics[width=0.8\linewidth]{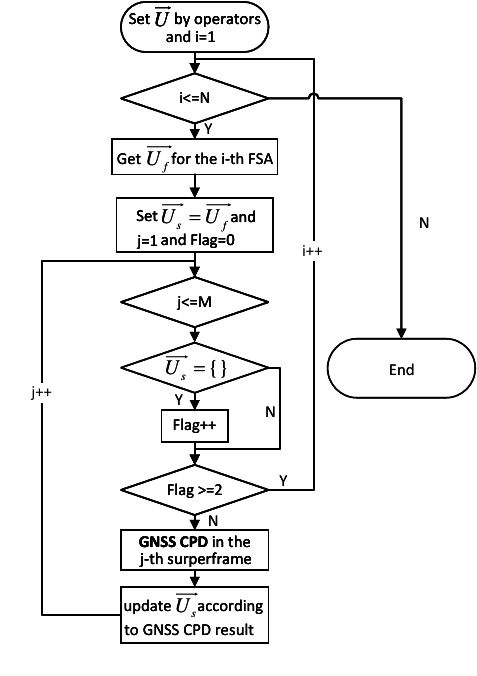}
		\caption{User Procedures: Flowchart illustrating the GNSS CPD.}
		\label{fig4}
	\end{figure}

	Figure~\ref{fig4} illustrates the GNSS user service procedure based on user requirements $\vec{U}$ over $N$ FSA periods, where each FSA consists of $M$ superframes. 
	The procedure ensures that GNSS schedules ISLs in alignment with the specified user requirements while adhering to its operational constraints and resource availability.
	
	\paragraph{Service Request on a FSA state ($\overrightarrow{{{U}_{f}}}$)}
	Represents the link establishment requests of users for a specific FSA. 
	Users within \(\vec{U}\) do not necessarily require ISLs in every FSA, leading to varying link establishment requests across FSAs. 
	For instance, given \(\vec{U} = \{(U_1, [2, 3, 4, 1])\}\), user \(U_1\) requests GNSS services every two FSAs. 
	\begin{itemize}
		\item \textbf{First FSA}: \(U_1\) has a link request, represented as \(\overrightarrow{{{U}_{f}}} = \{(U_1, [3, 4, 1])\}\). Here, \((U_1, [3, 4, 1])\) indicates that \(U_1\) requires ISLs with a duration of three slots for a total of four instances. At any given time, \(U_1\) can establish a link with only one GNSS satellite.
		\item \textbf{Second FSA}: \(U_1\) has no link request according to \(\vec{U}\), resulting in \(\overrightarrow{{{U}_{f}}} = \{\}\).
		\item \textbf{Third FSA}: \(U_1\) again has a link request, where \(\overrightarrow{{{U}_{f}}} = \{(U_1, [3, 4, 1])\}\).
	\end{itemize}
	This pattern repeats in subsequent FSAs. The described process aligns with the step 'Get \(\overrightarrow{{{U}_{f}}}\) for the \(i\)-th FSA' depicted in Fig.~\ref{fig4}.
	
	\paragraph{Service Request on a Superframe ($\overrightarrow{{{U}_{s}}}$)}
	Represents the link establishment requests of users under a specific superframe. 
	An FSA comprises multiple superframes, and the slot resources available within a single GNSS superframe are limited. 
	Consequently, fulfilling \(\overrightarrow{{{U}_{f}}}\) may require multiple superframes.
	In the first superframe of each FSA state, we initialize $\overrightarrow{{{U}_{s}}}$ to $\overrightarrow{{{U}_{f}}}$. Subsequently, $\overrightarrow{{{U}_{s}}}$ is updated in each following superframe based on the user service status of its preceding superframe.
	For example, let \(\overrightarrow{{{U}_{f}}} = \{(U_1, [3, 4, 1])\}\). Within the first superframe, \(\overrightarrow{{{U}_{s}}}\) is initialized as:
	\[
	\overrightarrow{{{U}_{s}}} = \overrightarrow{{{U}_{f}}} = \{(U_1, [3, 4, 1])\}.
	\]
	Assume that the GNSS CPD within the first superframe provides \(U_1\) with two ISLs of duration 3 slots each. The updated \(\overrightarrow{{{U}_{s}}}\) for the second superframe becomes:
	\[
	\overrightarrow{{{U}_{s}}} = \{(U_1, [3, 2, 1])\}.
	\]
	(This corresponds to 'update \(\overrightarrow{{{U}_{s}}}\) according to GNSS CPD result' as shown in Fig.~\ref{fig4}.)
	If the GNSS CPD provides \(U_1\) with two additional ISLs of duration 3 slots each within the second superframe, then the GNSS satisfies \(\overrightarrow{{{U}_{f}}}\) for this FSA during the second superframe. 
	Consequently, \(\overrightarrow{{{U}_{s}}}\) for the third superframe becomes:
	$\overrightarrow{{{U}_{s}}} = \{\}$.
	
	\paragraph{CPD Process State (Flag variable)}
	When $\overrightarrow{{{U}_{s}}} = \{\}$, it signifies that the GNSS no longer needs to provide links for users, prompting an increment in the Flag variable ($\text{Flag}++$). 
	The Flag variable represents the current state of the GNSS CPD process:
	\begin{enumerate}
		\item A Flag value of 0 indicates that the ongoing GNSS CPD must still establish user links.
		\item When the Flag equals 1, it indicates that the GNSS CPD only needs to consider internal ISLs within the GNSS constellation without providing ISLs to external users.
		\item If the Flag is greater than or equal to 2, it signifies that subsequent superframe CPD is unnecessary. The link relationships for the remaining superframes in the current FSA can directly utilize the GNSS CPD results obtained when Flag was 1.
	\end{enumerate}
	
	Continuing with the previous example, suppose in the third superframe $\overrightarrow{{{U}_{s}}} = \{\}$. 
	The Flag increments from 0 to 1, signifying that the GNSS CPD now exclusively plans ISLs within the GNSS constellation. 
	In the fourth superframe, the Flag increments from 1 to 2, exiting the CPD loop for the current FSA and proceeding to the next FSA. 
	By default, the GNSS CPD results for the fourth through the $M$-th superframe of the current FSA are assumed to be identical to the GNSS CPD results obtained when Flag was 1, corresponding to the third superframe in this example.
	
	\subsection{GNSS CPD Optimization Objective}
	
	\subsubsection{Maximize Ranging Diversity}
	
	GNSS employs two-way pseudorange ranging techniques. In this method, two satellites exchange ranging signals reciprocally: Satellite A transmits at time $t_1$ (recorded by A's clock), which is received by Satellite B at time $t_2$ (recorded by B's clock). Subsequently, Satellite B transmits at time $t_3$, which is received by Satellite A at time $t_4$. The four timestamps ($t_1$, $t_2$, $t_3$, $t_4$) are then used to compute the precise inter-satellite range and clock offset~\cite{25}.
	The PDOP (Position Dilution of Precision) metric is used to evaluate the ranging performance of GNSS~\cite{26}. 
	For a given satellite, an effective ranging link is defined as a link established with different satellites under a specified state. 
	With a fixed number of ranging links, the better the geometric configuration among the ranging satellites, the lower the PDOP value. 
	Moreover, it has been demonstrated in~\cite{27} that increasing the number of ranging links reduces the PDOP value. 
	Accordingly, GNSS satellites dynamically establish ISLs with multiple visible neighbors within a superframe. 
	The greater the diversity and quantity of ISLs, the richer the ranging information becomes, leading to a smaller PDOP value and more accurate OD for the source satellite.
	
	\subsubsection{Minimize Data Delivery Delay}
	
	Throughout its entire lifecycle, GNSS needs to transmit telemetry data to GSs. 
	A satellite visible to any GS is defined as an \textit{anchor satellite}, while those not visible are defined as \textit{non-anchor satellites}. 
	In this paper, we assume:
	\begin{enumerate}
		\item Satellites are equipped with dedicated satellite-to-ground links, allowing anchor satellites to directly transmit data to GSs;
		\item Non-anchor satellites implement a one-hop routing strategy, meaning they can only transmit data to anchor satellites, which then relay the data to GSs.
	\end{enumerate}
	Due to the evolving topology caused by GNSS CPD over time, ISLs between non-anchor and anchor satellites do not necessarily exist in every time slot. 
	When no link is available, data is temporarily stored on non-anchor satellites, resulting in a store-carry-forward flow~\cite{28,29} until the next connection opportunity arises between non-anchor and anchor satellites. 
	Given that a time slot is typically set to 3 seconds~\cite{30,27}, significantly longer than the propagation and transmission delays in GNSS (typically in the order of milliseconds), this paper defines the delay of a non-anchor satellite as the number of time slots it waits before relaying data to an anchor satellite, neglecting the minor propagation and transmission delays.
	
	\subsubsection{Maximize Data Throughput}
	The telemetry data from GNSS and the short message communication features available in some GNSS systems (e.g., BeiDou System) must be transmitted to GSs. 
	Given that anchor satellites can directly transmit data to GSs, the critical factor for the overall throughput of the GNSS system is the capability of non-anchor satellites to relay data to GSs. 
	Assuming that each ISL within GNSS has uniform transmission capacity, this paper defines the total number of ISLs established between non-anchor and anchor satellites as the throughput of the GNSS network.
	
	\subsubsection{Maximize User Satisfaction Ratio}
	Although the ISLs provided by GNSS to users serve different functions across various scenarios, this paper considers only the number of ISLs provided by GNSS to users without delving into the specific functionalities of these ISLs. 
	The user requirement satisfaction ratio for a given user is defined as the ratio of the number of ISLs provided by GNSS to the total number of link establishments required by the user, as specified in $\vec{U}$. 	
	GNSS CPD should aim to achieve a higher user requirement satisfaction ratio to better address user needs.
	
	It is important to note that if a user $U_n$ requests link establishment with a duration of $b_n$ time slots, but GNSS provides an ISL for fewer than $b_n$ slots, this ISL is considered substandard and is not counted toward the total number of ISLs provided by GNSS.
	To illustrate, consider a cislunar user requesting a 2-slot ISL from GNSS for navigation. If GNSS only provides a 1-slot ISL, this link is deemed inadequate. The reason is that, given the second-level propagation delay at the cislunar scale,  1-slot is insufficient to guarantee the completion of the two-way ranging between the GNSS and the user—meaning the ISL fails to serve its navigation function.
	
	The aforementioned CPD criteria represent unified yet conflicting objectives. 
	For instance, establishing links between non-anchor and anchor satellites may benefit ranging, communications, and throughput. 
	However, to meet the ranging needs of satellites, it may sometimes be necessary to establish ISLs between anchor satellites or between non-anchor satellites, which do not directly contribute to communication or throughput improvements.

	\section{ILP-Based GNSS CPD}
	\label{sec_ilp_based_gnss_cpd}
	
	In this paper, we utilize a general GNSS scenario comprising $N_s$ satellite nodes, denoted as $V_s$, and $N_u$ user nodes, denoted by $V_u$. 
	It should be noted that $V_u$ encompasses only the users included in $\overrightarrow{{{U}_{s}}}$, where $\overrightarrow{{{U}_{s}}}$ denotes the user requests within the current superframe. 
	$V_u$ varies from one superframe to another.
	The entire set of nodes within this network is represented as $V=V_s \cup V_u=\{v_i|1\leq i \leq N\}$, where $N$ is the total number of nodes and $N=N_s+N_u$. 
	Subsequently, the set of satellite nodes is categorized into anchor and non-anchor satellites, expressed as $V_s=V_a \cup V_n$. Let $N_a$ and $N_n$ denote the number of anchor and non-anchor satellites, respectively. 
	Consequently, we have $N=N_a+N_n+N_u$.
	
	For simplicity, we model the problem as an ILP using a single superframe under a given FSA state (corresponds to 'GNSS CPD in the j-th superframe' as shown in Fig.~\ref{fig4}). 
	For this specific FSA, we introduce a 0-1 visibility matrix $Y$, and for a particular superframe within that FSA, we introduce a 0-1 matrix $X$ corresponding to the ISL relationships among nodes as derived from GNSS CPD.
	As previously mentioned, to gather more ranging information, a superframe is subdivided into several equal-length time slots. 
	We index the slots within a single superframe by $k\in T=\{1, 2, ..., K\}$. 
	In matrix $X$, $x_{i,j,k}=1$ represents that nodes $v_i$ and $v_j$ establish an ISL during the $k$-th time slot of the superframe, while $x_{i,j,k}=0$ indicates no ISL is established. 
	In matrix $Y$, $y_{i,j}=1$ indicates $v_i$ and $v_j$ are mutually visible~(i.e., capable of establishing an ISL) within the specified FSA state, whereas $y_{i,j}=0$ indicates they cannot establish an ISL.
	
	\subsection{Preprocessing}
	
	Before formulating the ILP model, preprocessing of the visibility matrix $Y$ is required. 
	We set all visibility between users to 0, thereby preventing link establishment directly between users. 
	For constellation users, the link establishment among nodes within the constellation is determined by the constellation CPD rather than the GNSS CPD.
	
	\subsection{ILP Model}
	\subsubsection{General Constraints}
	The following are three general constraints concerning link relationships matrices $X$ and visibility matrices $Y$:
	\begin{equation}
		x_{i,j,k}=\{0,1\}, \forall v_i,v_j \in V, k\in T,\label{eq2}
	\end{equation}
	\begin{equation}
		x_{i,j,k}=x_{j,i,k}, \forall v_i,v_j \in V,  k\in T,\label{eq3}
	\end{equation}
	and 
	\begin{equation}
		x_{i,j,k}\leq y_{i,j}, \forall v_i,v_j \in V,  k\in T,\label{eq4}
	\end{equation}
	where \eqref{eq3} represents that the ISL established between $v_i$ and $v_j$ is bidirectional. 
	\eqref{eq4} indicates that an ISL can only be established between $v_i$ and $v_j$ if they are mutually visible.
	It should be noted that, given the ample link budget between GNSS satellites and Cislunar users computed in Section~\ref{sec_background}-A-1-c, this paper does not consider scenarios where a link, although visible, cannot be established due to insufficient link budget.
	
	Additionally, given that each GNSS satellite is equipped with only one ISL transponder and user $v_i$ carries $d_i$ ISL transponders, we have:
	\begin{equation}
		\sum_{v_i\in V} x_{i,j,k} \leq 1,  \forall v_j \in V_s,  k\in T,\label{eq5}
	\end{equation}
	\begin{equation}
		\begin{split}
			\sum_{v_i\in V} x_{i,j,k} \leq d_i,  \forall v_j \in V_u,  k\in T, \\(v_j,[b_j,c_j,d_j])\in \overrightarrow{{{U}_{s}}}.\label{eq6}
		\end{split}
	\end{equation}
	\eqref{eq5} indicates that for a specific GNSS satellite, only one ISL can be established with one entity within a time slot. 
	\eqref{eq6} ensures that for a specific user $v_i$, it can establish up to $d_i$ ISLs with $d_i$ different entities within a time slot.
	
	\subsubsection{Ranging Constraints}
	Given that the PDOP is nonlinear and challenging to express as linear constraints, it has been proven in~\cite{27} that an increase in the number of ranging links correlates with a decrease in PDOP values. 
	Therefore, this paper adopts the number of ranging links as a metric for assessing ranging performance. 
	By enforcing the constraints on the number of ranging links as detailed in~\cite{31,17}, we assume that as long as all satellites establish more than $L_{min}$ ranging links, sufficient ranging information can be obtained to achieve high-precision OD~\cite{32}. 
	We have:
	\begin{equation}
		l_{i,j}=\left\{ \begin{matrix} 1, &\sum_{ k\in T} x_{i,j,k}\geq 1, \forall v_i,v_j \in V_s \\ 0, &\sum_{ k\in T} x_{i,j,k}=0 , \forall v_i,v_j \in V_s &\\ \end{matrix} \right. \label{eq7}
	\end{equation}
	\begin{equation}
		\sum_{ v_j\in V_s}l_{i,j} \geq L_{min}, \forall v_i \in V_s .\label{eq8}
	\end{equation}
	Wherein $l_{i,j}$ represents the presence or absence of an ISL between GNSS satellite nodes $v_i$ and $v_j$ within the $K$ slots of the superframe. 
	Repeated link establishments between $v_i$ and $v_j$ across multiple slots within the same superframe do not contribute to ranging performance.
	Since \eqref{eq8} is not a linear expression, the Big-M method is employed to transform it into a linear form:
	\begin{equation}
		l_{i,j}=\{0,1\}, \forall v_i,v_j \in V_s ,\label{eq9}
	\end{equation}
	\begin{equation}
		l_{i,j}\leq \sum_{ k\in T}x_{i,j,k} \leq Ml_{i,j},  \forall v_i,v_j \in V_s .\label{eq10}
	\end{equation}
	Here, M is a constant chosen to be greater than the sum of $x_{i,j,k}$. If $\sum_{ k\in T}x_{i,j,k}=0$, then to satisfy $l_{i,j}\leq \sum_{ k\in T}x_{i,j,k}$, it follows that $l_{i,j}=0$. Conversely, if  $\sum_{ k\in T}x_{i,j,k} \geq 1$, to satisfy $\sum_{ k\in T}x_{i,j,k} \leq Ml_{i,j}$, it must be that $l_{i,j}=1$ . 
	
	\subsubsection{Communication Constraints}
	This paper assumes satellites have dedicated satellite-to-ground links, enabling anchor satellites to transmit data directly to GSs. 
	Consequently, our focus is solely on modeling the communication constraints for non-anchor satellites. 
	We employ the method for addressing communication constraints proposed in our previous work~\cite{17}.
	
	In compliance with the one-hop routing strategy stated in Sec~\ref{sec_system_model}-C, all non-anchor satellites aim to establish ISLs with anchor satellites frequently and uniformly. 
	This ensures they can forward data in available slots to anchor satellites quickly. 
	Toward this objective, we introduce a 2-dimensional matrix $\psi$ with $N_n$ rows and $K$ columns to represent the connectivity of non-anchor satellites to anchor satellites in a given slot, which is expressed as:
	\begin{equation}
		\psi_{i,k} =\sum_{v_j \in V_a}x_{i,j,k}, \forall v_i \in  V_n, k\in T.\label{eq11}
	\end{equation}
	From \eqref{eq5}, it is evident that $\psi_{i,k}$ is a Boolean variable. 
	Specifically, $\psi_{i,k}=1$ indicates that the i-th non-anchor satellite is connected to an anchor satellite in the k-th slot, thereby establishing a communication ISL, and $\psi_{i,k}=0$ otherwise. 
	To minimize the maximum delay of data generated by any non-anchor satellite within any slot, we introduce $T_m$ and construct a matrix $A$ as follows:
	\begin{equation}
		\begin{array}{l}
			\hspace*{1.15cm}\overbrace{\hspace*{1.7cm}}^{T_m} \hspace*{0.5cm} \overbrace{\hspace*{2.5cm}}^{(K-T_m)}\\ [-1ex]
			\boldsymbol{A}=\left[\begin{array}{ccc}
				1 & \cdots & 1 \\
				0 & 1 & \cdots \\
				\vdots & \vdots & \ddots \\
				0 & \cdots & \cdots
			\end{array}\right.
			\left.\left.\begin{array}{cccc}
				0 & \cdots & \cdots & 0 \\
				1 & 0 & \cdots & 0 \\
				\vdots & \vdots & \ddots & \vdots \\
				0 & 1 & \cdots & 1
			\end{array}\right]\right\}\left(K-T_m+1\right) \label{eq12}
		\end{array}
	\end{equation}
	Therefore, the communication constraints can be expressed as follows:
	\begin{equation}
		\begin{array}{l}
			\hspace*{1.6cm}\overbrace{\hspace*{1.6cm}}^{K-T_m+1} \hspace*{0.5cm}\\ [-1ex]
			\psi A^{T} \geq 
			\left.\left[\begin{array}{ccc}
				1 & \cdots & 1  \\
				\vdots & \ddots & \vdots \\
				1 & \cdots & 1  \\
			\end{array}\right]\right \} N_n
		\end{array} \label{eq13}
	\end{equation}
	Specifically, suppose the i-th non-anchor satellite does not establish ISL to an anchor satellite for $T_m$ or more consecutive slots. 
	In that case, there will be at least one 0 in the i-th row of the left-hand side matrix $\psi A^{T}$, thereby violating the constraint. 
	In other words, the constraint ensures that the maximum delay for non-anchor satellites is less than $T_m$.
	
	\subsubsection{User Requirements Constraints}
	The following constraints aim to provide the user with ISLs according to the user requirement $\overrightarrow{{{U}_{s}}}$ within this superframe on a best-effort basis.
	In the following equation:
	\begin{equation}
		\begin{split}
			r_{i,j,k}=\{0,1\} , \forall v_i \in V_u~and~(v_i,[b_i,c_i,d_i])\in \overrightarrow{{{U}_{s}}},
			\\\forall v_j \in V_s, 1 \leq k \leq K-b_i+1,\label{eq14}
		\end{split}
	\end{equation}
	$r_{i,j,k}$ represents whether user $v_i$ establishes a continuous ISL  with GNSS satellite $v_j$ starting from k-th time slot for $b_i$ consecutive slots. $r_{i,j,k}=1$ indicating a successful ISL for user $v_i$.
	To determine the value of $r_{i,j,k}$ we use the following equations based on the link establishment relationship between user $v_i$ and satellite $v_j$ over $b_i$ consecutive slots, where the sum of $k$ and $dk$ does not exceed $K$:
	\begin{equation}
		\begin{split}
			r_{i,j,k} \leq x_{i,j,k+dk} , 
			\\ \forall v_i \in V_u~and~(v_i,[b_i,c_i,d_i])\in \overrightarrow{{{U}_{s}}},
			\\ \forall v_j \in V_s, 1 \leq k \leq K-b_i+1, 0 \leq dk \leq b_i-1. \label{eq15}
		\end{split}
	\end{equation}

	\begin{equation}
		\begin{split}
			r_{i,j,k} \geq \sum_{0 \leq dk \leq b_i-1}x_{i,j,k+dk}-b_i+1 ,  
			\\ \forall v_i \in V_u~and~(v_i,[b_i,c_i,d_i])\in \overrightarrow{{{U}_{s}}},
			\\ \forall v_j \in V_s, 1 \leq k \leq K-b_i+1. \label{eq16}
		\end{split}
	\end{equation}
	
	For user $v_i$ and satellite $v_j$, starting from the $k$-th time slot, if any $x_{i,j,k+dk}$ within the consecutive $b_i$ slots is 0, \eqref{eq15} necessarily imposes the constraint $r_{i,j,k} \leq 0$, and \eqref{eq16} imposes the constraint $r_{i,j,k} \geq \text{a non-positive integer}$. 
	Combining \eqref{eq15}, \eqref{eq16}, and \eqref{eq14}, $r_{i,j,k}$ can only take the value of 0.
	
	Conversely, for user $v_i$ and satellite $v_j$, starting from the $k$-th time slot, if all $x_{i,j,k+dk}$ within the consecutive $b_i$ slots are 1, then \eqref{eq15} imposes the constraint $r_{i,j,k} \leq 1$, and \eqref{eq16} imposes the constraint $r_{i,j,k} \geq 1$. 
	Combining \eqref{eq15}, \eqref{eq16}, and \eqref{eq14}, $r_{i,j,k}$ can only take the value of 1.
	
	To prevent satellite $v_j$ from providing an ISL to user $v_i$ that exceeds $b_i$ slots, we use the following constraint:
	\begin{equation}
		\begin{split}
			r_{i,j,k}+r_{i,j,k+1} \leq 1,  
			\\ \forall v_i \in V_u~and~(v_i,[b_i,c_i,d_i])\in \overrightarrow{{{U}_{s}}},
			\\ \forall v_j \in V_s, 2 \leq k+1 \leq K-b_i+1.
			\label{eq17}
		\end{split}
	\end{equation}
	For instance, if $b_i$=2, and starting from the k-th time slot, satellite $v_j$ provides an ISL to user $v_i$ for a duration of 3 slots, this would result in both $r_{i,j,k}$ and $r_{i,j,k+1}$ being 1. 
	This situation could be mistakenly interpreted as satellite $v_j$ providing two separate ISLs, each lasting $b_i$ slots, to user $v_i$. 
	\eqref{eq17} avoids such a scenario, ensuring that the ISL duration does not surpass the specified limit of $b_i$ slots.
	
	Given that the slot resources of GNSS within a superframe are limited, the GNSS CPD in one superframe may not be able to provide user $v_i$ with $c_i$ instances of ISLs each lasting $b_i$ slots. 
	If we directly apply $= c_i$ as a constraint, it might lead to an infeasible solution. 
	Therefore, we introduce $p_i$: 
	\begin{equation}
		\begin{split}
			p_i \in \mathbb{N}, \forall v_i \in V_u~and~(v_i,[b_i,c_i,d_i])\in \overrightarrow{{{U}_{s}}},
			\label{eq18}
		\end{split}
	\end{equation}
	where $\mathbb{N}$ denotes the set of natural numbers, and each user $v_i$ corresponds to a $p_i$. 
	To specify the number of link establishments provided for user $v_i$ should be equal to $c_i-p_i$ we use:
	\begin{equation}
		\begin{split}
			\sum_{v_j \in V_s} \sum_{~1\leq k \leq K-b_i+1} r_{i,j,k} = c_i-p_i,  
			\\ \forall v_i \in V_u~and~(v_i,[b_i,c_i,d_i])\in \overrightarrow{{{U}_{s}}}.
			\label{eq19}
		\end{split}
	\end{equation}
	This parameter indicates that it is not mandatory to provide the complete $c_i$ services within one superframe; instead, only $c_i-p_i$ services need to be delivered within the current superframe. 
	The remaining $p_i$ services are deferred and stored in the next superframe's $\overrightarrow{{{U}_{s}}}$, to be satisfied by the GNSS CPD of the subsequent superframe.
	
	In contrast to communication and ranging constraints, considered hard constraints, user requirement constraints are classified as soft constraints. 
	The ILP methodology inherently allows flexibility in addressing partially unmet user requirements.
	
	\subsubsection{Objective Function}
	According to the constraints above, the problem of topology design in GNSSs falls in the category of 0-1 ILP as follows:
	\begin{equation}
		\begin{split}
			max ~\sum_{v_i \in V_n} \sum_{~1\leq k \leq K}  \psi_{i,k} - C*\sum_{v_j \in V_u} p_j
			\\s.t.~~~~~~~ \eqref{eq2}-\eqref{eq6}, \eqref{eq8}-\eqref{eq19}.
			\label{eq20}
		\end{split}
	\end{equation}
	which can be solved by using some off-the-shelf solvers, such as Gurobi.
	In \eqref{eq20}, $C$ is a configurable constant. 
	Here, the summation $\sum_{v_i \in V_n} \sum_{~1\leq k \leq K}  \psi_{i,k}$ describes the total number of ISLs between non-anchor satellites and anchor satellites. 
	A larger value of this summation indicates a higher throughput and, consequently, a larger objective function value. 
	The term $C*\sum_{v_j \in V_u} p_j$ serves as a penalty for unmet user requests by the GNSS CPD; the more ISLs the GNSS CPD fails to provide, the larger the penalty, leading to a smaller objective function value.
	
	Overall, the ILP-based GNSS CPD is designed to ensure the availability of essential resources for the GNSS constellation's operation, including communication and ranging constraints. 
	The CPD scheme also aims to optimize the constellation's operational performance, reflected in the throughput component of the objective function, as well as enhance its external service capabilities, reflected in the penalty part of the objective function.
	
	\subsection{Postprocessing}
	
	\subsubsection{Update $\overrightarrow{{{U}_{s}}}$ for the subsequent superframe}
	
	This part corresponds to 'update $\overrightarrow{{{U}_{s}}}$ according to GNSS CPD result' as shown in Fig.~\ref{fig4}.
	The $\overrightarrow{{{U}_{s}}}$ for the next superframe (if the current FSA has a subsequent superframe) can be determined based on the $p_i$ values derived from the CPD outcome of the current superframe.
	
	Assume that within the current superframe, $\overrightarrow{{{U}_{s}}}=\{(v_1,[2,3,1],v_2,[1,4,1])\}$, where $(v_1, [2, 3, 1])$ signifies that user $v_1$ requests the GNSS to establish an ISL lasting for 2 slots, repeated three times, and $v_1$ is equipped with a single terminal. 
	Similarly,  $(v_2, [1, 4, 1])$ indicates that user $v_2$ requests the GNSS to establish an ISL lasting for 1 slot, repeated four times, and $v_2$ also possesses a single terminal.
	Given the CPD results, we have $p_1=1$ and $p_2=0$. The value $p_1=1$ indicates that GNSS still needs to provide one ISL for $v_1$ in the subsequent superframe, while $p_2=0$ indicates that no further ISLs are requeired for $v_2$.
	Consequently, it is straightforward to determine that $\overrightarrow{{{U}_{s}}}$ for the next superframe will be $\overrightarrow{{{U}_{s}}}=\{(v_1,[2,1,1])\}$. 
	
	\subsubsection{Remove substandard ISLs provided to users}
	
	In the results obtained from GNSS CPD, ISLs may be provided to users who do not fulfill the specified duration requirements. 
	For instance, with $\overrightarrow{{{U}_{s}}}=\{(U_1,[3,3,1])\}$, GNSS might provide ISLs to $U_1$ with durations of one slot or two slots, which are considered substandard for $U_1$. 
	These substandard ISLs are generated under the constraints of communication and ranging while not benefiting the GNSS throughput. 
	To conserve on-board resources of GNSS satellites, these substandard ISLs need to be removed.
	The process is presented as Algorithm~\ref{alg1}, where \textit{StandardIndex} stores the indices of standard ISLs.
	
	\begin{algorithm}
		\caption{Remove substandard ISLs for users}
		\label{alg1}
		\begin{algorithmic}[1]
			\Require $V_u$, $V_s$, $T$ , $\overrightarrow{{{U}_{s}}}$, GNSS CPD result (i.e., $x, r$)
			\Ensure GNSS CPD result $x$ without substandard ISLs
			\State \textbf{Begin}
			\State StandardIndex=[]
			\For {$v_i \in V_u~and~(v_i,[b_i,c_i,d_i])\in \overrightarrow{{{U}_{s}}}$}
			\For{$v_j \in V_s$}
			\For{$1\leq k\leq K-b_i+1$}
			\If{$r_{i,j,k}$==1}
			{
				\For{$0\leq dk\leq b_i-1$}
				\\~~~~~~~~~~~~~~~~~~~~~StandardIndex.add($(i, j, k+dk)$)
				\EndFor
			}
			\EndIf
			\EndFor
			\EndFor
			\EndFor
			\For {$v_i \in V_u~and~(v_i,[b_i,c_i,d_i])\in \overrightarrow{{{U}_{s}}}$}
			\For{$v_j \in V_s$}
			\For{$1\leq k\leq K$}
			\If {$(i,j,k)$ $\notin$ StandardIndex}
			{
				\\~~~~~~~~~~~~~~~~~$x_{i,j,k}=0$,
				\\~~~~~~~~~~~~~~~~~$x_{j,i,k}=0$,
			}
			\EndIf
			\EndFor
			\EndFor
			\EndFor
		\end{algorithmic}
	\end{algorithm}

	\section{Evaluation}
	\label{sec_evaluation}
	
	This section is structured into three parts. 
	The first part analyzes the parameters of the ILP-based GNSS CPD (Section~\ref{eval_part1}). 
	The second part evaluates the performance of the proposed ILP-based GNSS CPD (Section~\ref{eval_part2}). 
	The third part analyzes the BeiDou ISL use case on the OD accuracy of Earth-Moon LP constellations (Section~\ref{eval_part3}).
	All experiments were conducted on a workstation equipped with an AMD Ryzen 7 7700X processor (4.5 GHz) and 32 GB of RAM. The ILP problems were solved using the Gurobi solver, which was invoked within the PyCharm using Python 3.12.3.
	
	\paragraph{Considerations}
	As a general consideration, it is important to note that in this paper:
	\begin{itemize}
		\item \textbf{Throughput} refers to the number of ISLs established between non-anchor and anchor satellites.
		\item \textbf{Delay} refers to the number of time slots a non-anchor satellite must wait to establish an ISL with an anchor satellite.
	\end{itemize}
	Finally, since the duration of a time slot far exceeds the propagation and transmission delays, these latter delays are neglected in this study~\cite{27}.
	
	Previous studies~\cite{17,46} commonly assume that only one superframe’s link scheduling result needs to be calculated for a given FSA state, with the remaining superframes within the FSA repeating this result. 
	In contrast, this paper addresses the extended applications of GNSS ISLs, where the number of superframes requiring calculation within an FSA is contingent on user link establishment requests. 
	Once all user demands are fulfilled, a standalone GNSS CPD is performed without considering ISL extensions. 
	The results of this GNSS CPD are then reused for the remaining superframes in the FSA.
	
	For example, suppose all user link establishment requests are satisfied in the first superframe. 
	In that case, the second superframe proceeds with GNSS CPD for internal operations, and the scheduling results of the second superframe are repeated across the remaining superframes.
	
	This approach effectively balances the requirements of extended GNSS ISL applications while optimizing onboard memory usage. 
	
	\paragraph{Scenario}
	We utilize the BeiDou system as the exemplary GNSS constellation, comprising 24 MEO satellites, 3 GEO satellites, 3 IGSO satellites, and three GSs. 
	The orbital parameters and GS coordinates are detailed in Table~\ref{Orbit parameters and GS locations}. 
	Additional fundamental parameters employed in the simulation are likewise tabulated in Table~\ref{Basic parameters in the simulation}. 
	We define an FSA duration as 5 minutes, with each FSA encompassing five superframes.

	\begin{table}[]
		\caption{Orbit parameters and ground station locations}
		\label{Orbit parameters and GS locations}
		\centering
		\footnotesize
		\renewcommand{\arraystretch}{1.2} 
		\begin{tabular}{|c|c|}
			\hline
			\textbf{Satellites and GSs} & \textbf{Description}  \\
			\hline
			MEO & Walker-$\delta$ 24/3/1, $h=21528$ km, $i=55^\circ$   \\ 
			\hline
			IGSO & $h=35786$ km, $i=55^\circ$, interval = $120^\circ$  \\
			\hline
			GEO & $h=35786$ km, lon = ($80^\circ$, $110.5^\circ$, $140^\circ$)   \\
			\hline
			Jiamusi & ($46.8^\circ$N, $130.3^\circ$E) \\
			\hline
			Kashi & ($39.47^\circ$N, $75.99^\circ$E)\\
			\hline
			Sanya & ($18.23^\circ$N, $109.02^\circ$E)\\
			\hline
		\end{tabular}
	\end{table}
	
	\begin{table}[]
		\caption{Basic parameters in the simulation}
		\label{Basic parameters in the simulation}
		\centering
		\footnotesize
		\renewcommand{\arraystretch}{1.2} 
		\begin{tabular}{|c|c|}
			\hline
			\textbf{Parameter} & \textbf{Value}   \\ 
			\hline
			Length of an FSA state & 5 min   \\ 
			\hline
			Length of a superframe & 1 min  \\
			\hline
			Length of a time slot & 3 s   \\
			\hline
			ISL pointing range in MEO & $60^\circ$ \\
			\hline
			ISL pointing range in GEO/IGSO & $45^\circ$ \\
			\hline
			GS pointing range & $85^\circ$ \\
			\hline
		\end{tabular}
	\end{table}

	\subsection{ILP Parameter Tuning}
	\label{eval_part1}
	
	This section discusses the key parameters used in the ILP model.
	
\begin{figure}[] 
		\centering
		\includegraphics[width=0.85\linewidth]{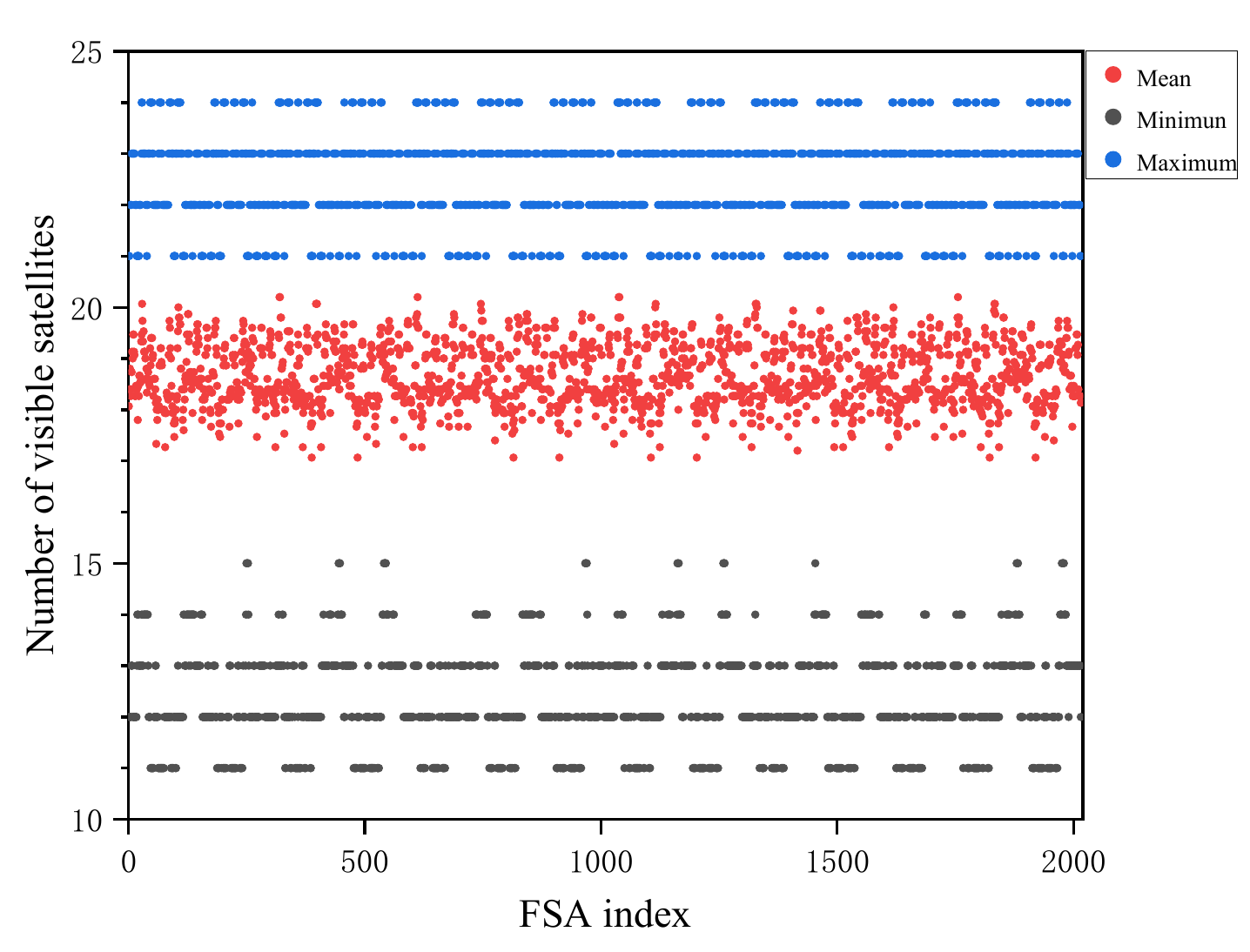}
		\caption{Number of visible satellites for satellites in all FSA states.}
		\label{fig5}
	\end{figure}

	The number of ranging links for GNSS satellites is directly proportional to OD accuracy~\cite{27}. Thus, the lower bound constraint on the number of GNSS satellite ranging links, denoted as $L_{min}$, should be set as high as feasible to optimize performance.
	Figure~\ref{fig5} illustrates the minimum, average, and maximum number of visible satellites for any GNSS satellite over a seven-day period (comprising 2016 FSA states), which corresponds to one regression cycle of the BeiDou constellation. 
	The data reveals that the minimum number of visible satellites is 11, indicating that some satellites can establish links with a maximum of 11 other satellites. Based on this observation, we set $L_{min} = 11$ for this study.
	
	Our previous work~\cite{17} analyzed the distribution of non-anchor and anchor satellites, concluding that a non-anchor satellite should ideally establish a link with an anchor satellite within at most one time slot, leading to a recommended $T_m$ value of 2. 
	However, this paper focuses on extended ISL application scenarios, particularly those involving Earth-Moon LP users. 
	For these scenarios, the propagation delay for links between GNSS satellites and Earth-Moon LP users is on the order of seconds. 
	To accommodate two-way pseudorange ranging, GNSS must provide ISLs with a minimum duration of two time slots (6 seconds) for these users.	
	If $T_m$ were set to 2, providing continuous ISLs spanning two time slots for Earth-Moon LP users by non-anchor satellites would conflict with the delay constraints. 
	To address this, we have set $T_m = 3$, ensuring that non-anchor satellites retain the flexibility to serve Earth-Moon LP users effectively while maintaining compliance with delay requirements.
	
	\begin{table}[]
		\caption{ILP Parameters Setup}
		\label{ILP parameters setup}
		\centering
		\footnotesize
		\renewcommand{\arraystretch}{1.2} 
		\begin{tabular}{|c|c|c|}
			\hline
			\thead{Parameter} & \thead{Description} & \thead{Value} \\ 
			\hline
			$L_{min}$ & Minimum number of ranging links & 11 \\ 
			\hline
			$T_m$ & Delay constraint for non-anchor satellites & 3 \\ 
			\hline
			$C$ & \makecell{Penalty factor for unmet user demands \\in the  ILP optimization objective}& 1000 \\ 
			\hline
			$M$ & Big $M$ parameter for constraints & 30 \\ 
			\hline
		\end{tabular}
	\end{table}
	
	Furthermore, given that the constellation throughput can already be maintained at a high level under the delay constraints of non-anchor satellites, we set $C = 1000$ in the optimization objective. 
	This configuration ensures that, during the ILP-solving process, if a conflict arises between maximizing constellation throughput and meeting user demands, the solution prioritizes satisfying user demands as deemed optimal.
	
	Lastly, the value of the big M parameter was set to 30, which proved sufficient for our simulation purposes.
	
	The values of the aforementioned parameters are summarized in Table~\ref{ILP parameters setup}. 
	All subsequent simulation settings are based on these parameter configurations.

	\subsection{ILP Performance Evaluation}
	\label{eval_part2}
	
	This section analyzes the performance of the ILP-based GNSS CPD. 
	We first evaluate the extended capacity of GNSS ISLs within a single superframe, followed by an examination of GNSS performance over one complete regression cycle.
	
	\subsubsection{Single Superframe}
	
	Table~\ref{anchor-satellites-states-statistics} summarizes the number of anchor satellites and state occurrences over one regression cycle of GNSS. To analyze the external expansion capacity of GNSS ISLs within a superframe, we focus on three representative states: 
	\begin{enumerate}
		\item The state with the minimum number of anchor satellites (12).
		\item The state with the maximum number of anchor satellites (19).
		\item The state that occurs most frequently (16).
	\end{enumerate}
	
	\begin{table}[tbp]
		\caption{Number of Anchor Satellites and States Statistics}
		\label{anchor-satellites-states-statistics}
		\centering
		\footnotesize 
		\renewcommand{\arraystretch}{1.2} 
		\resizebox{0.48\textwidth}{!}{
		\begin{tabular}{|c|c|c|c|c|c|c|c|c|}
			\hline
			\thead{Number of \\ Anchor Satellites} & 12 & 13 & 14 & 15 & 16 & 17 & 18 & 19 \\
			\hline
			\thead{Number of \\ States}            & 10 & 41 & 103 & 365 & 707 & 634 & 154 & 2 \\
			\hline
		\end{tabular}}
	\end{table}
	
	We define the \textbf{external extended capacity} of a superframe under a specific FSA as follows:
	If adding any additional link establishment request $c_n$ to any element $(U_n, [b_n, c_n, d_n])$ in the $\overrightarrow{{{U}_{s}}}$ corresponding to this superframe results in a non-zero penalty term $\sum_{v_j \in V_u} p_j$ during the ILP solving process, such a $\overrightarrow{{{U}_{s}}}$ is considered the external extended capacity of this superframe.
	
	The definition of expansion capacity is inherently linked to the users included in $\overrightarrow{{{U}_{s}}}$. 
	For different users, the expansion capacity of a superframe varies, depending on the visibility between the users and the overall GNSS constellation. 
	Therefore, the expansion capacity precisely reflects the maximum link establishment capability that the superframe can provide to each user in $\overrightarrow{{{U}_{s}}}$.
	
	We conducted expansion capacity tests under two case scenarios. 
	The first test case included two GEO users and two IGSO users, while the second test case comprised one L3 point user, one L4 point user, one L5 point user, and one lunar distant retrograde orbit (DRO) user. 
	Considering the propagation delay, GEO and IGSO users only need to request ISLs with a duration of one time slot, whereas LP users require ISLs with a duration of two time slots.
	
	The results obtained from these tests are summarized in Table~\ref{GNSS service capability}. 
	Here, $U_{GEO_1}$ represents the first user located in the GEO orbit, and the rest are denoted similarly.
	\begin{table}[]
		\caption{GNSS Service Capability}
		\label{GNSS service capability}
		\centering
		\footnotesize 
		\renewcommand{\arraystretch}{1.2} 
		\resizebox{0.49\textwidth}{!}{
		\begin{tabular}{|c|c|c|}
			\hline
			\thead{Number of \\ Satellites} & \thead{Capacity in the \\ First Test Case} & \thead{Capacity in the \\ Second Test Case} \\
			\hline
			\thead{12, 16, \\ or 19} & 
			\thead{
				$\overrightarrow{{{U}_{s1}}}$=\{($U_{GEO_1}$, [1, 20, 1]), \\
				($U_{GEO_2}$, [1, 20, 1]), \\
				($U_{IGSO_1}$, [1, 20, 1]), \\
				($U_{IGSO_2}$, [1, 20, 1])\}
			} & 
			\thead{
				$\overrightarrow{{{U}_{s2}}}$=\{($U_{L3_1}$, [2, 10, 1]), \\
				($U_{L4_1}$, [2, 10, 1]), \\
				($U_{L5_1}$, [2, 10, 1]), \\
				($U_{DRO_1}$, [2, 10, 1])\}
			} \\
			\hline
		\end{tabular}}
	\end{table}	
	In Table~\ref{GNSS service capability}, the expansion capacity exhibited in different states is identical regardless of the number of anchor satellites. 
	This is because, when both anchor and non-anchor satellites can serve users, the status of GNSS relative to GSs is transparent to the users. 
	If we increase the delay constraints such that non-anchor satellites cannot provide multi-slot ISLs to users, the expansion capacity would then differ depending on the number of anchor satellites.
	
\begin{figure}[] 
		\centering
		\includegraphics[width=0.8\linewidth]{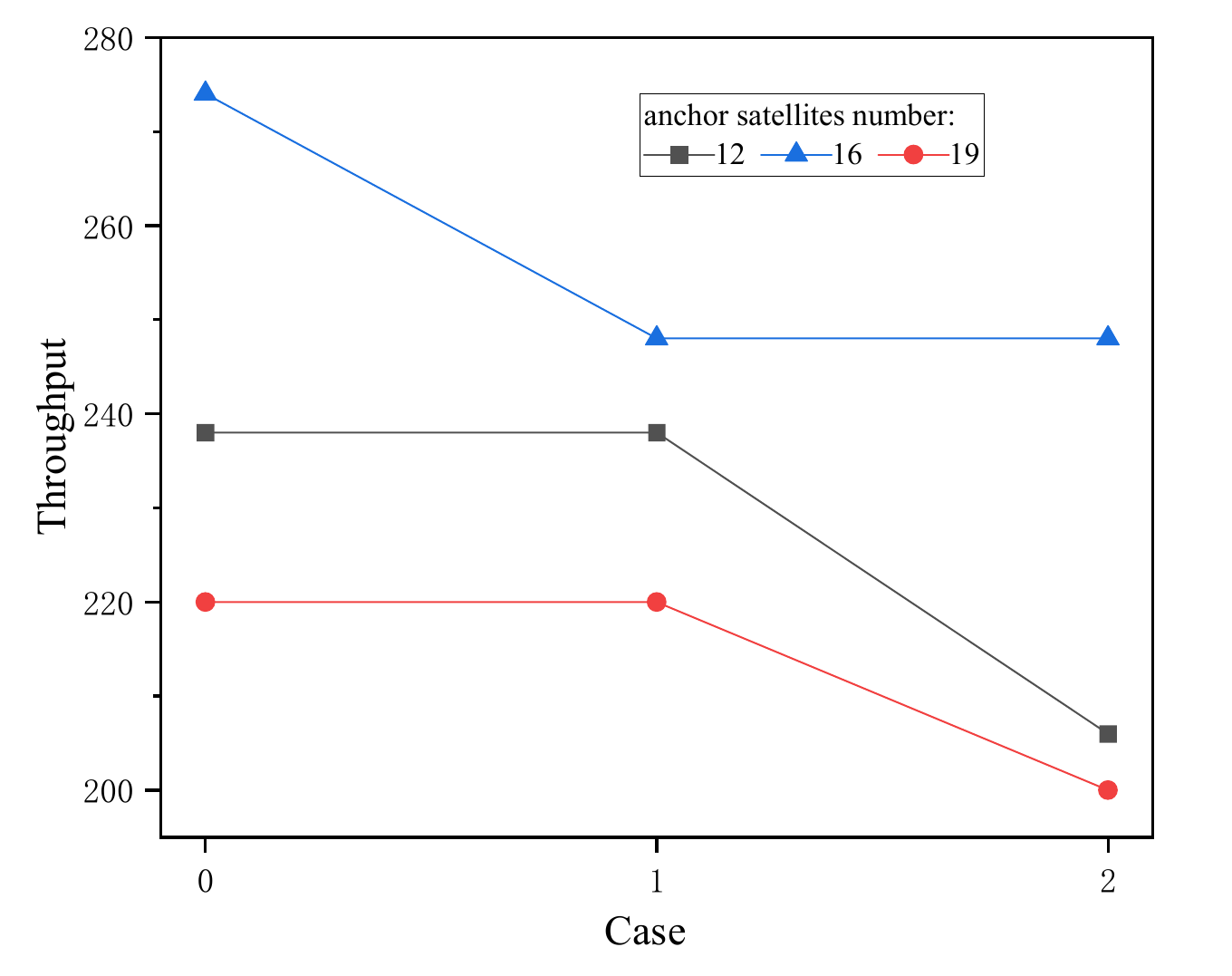}
		\caption{Throughput in a single superframe.}
		\label{fig6}
	\end{figure}

\begin{figure}[] 
		\centering
		\includegraphics[width=0.8\linewidth]{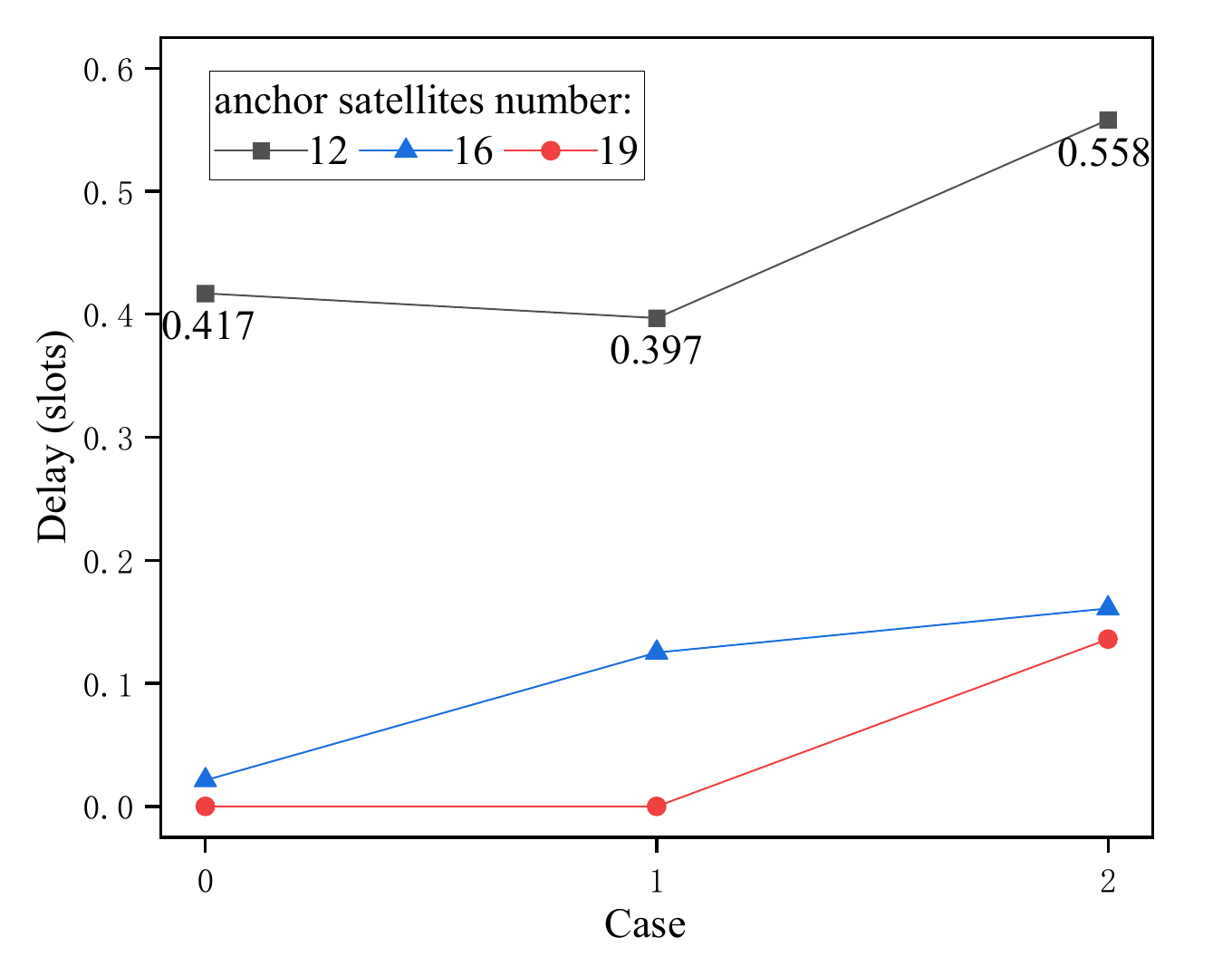}
		\caption{Delay in a single superframe.}
		\label{fig7}
	\end{figure}

\begin{figure}[] 
		\centering
		\includegraphics[width=0.8\linewidth]{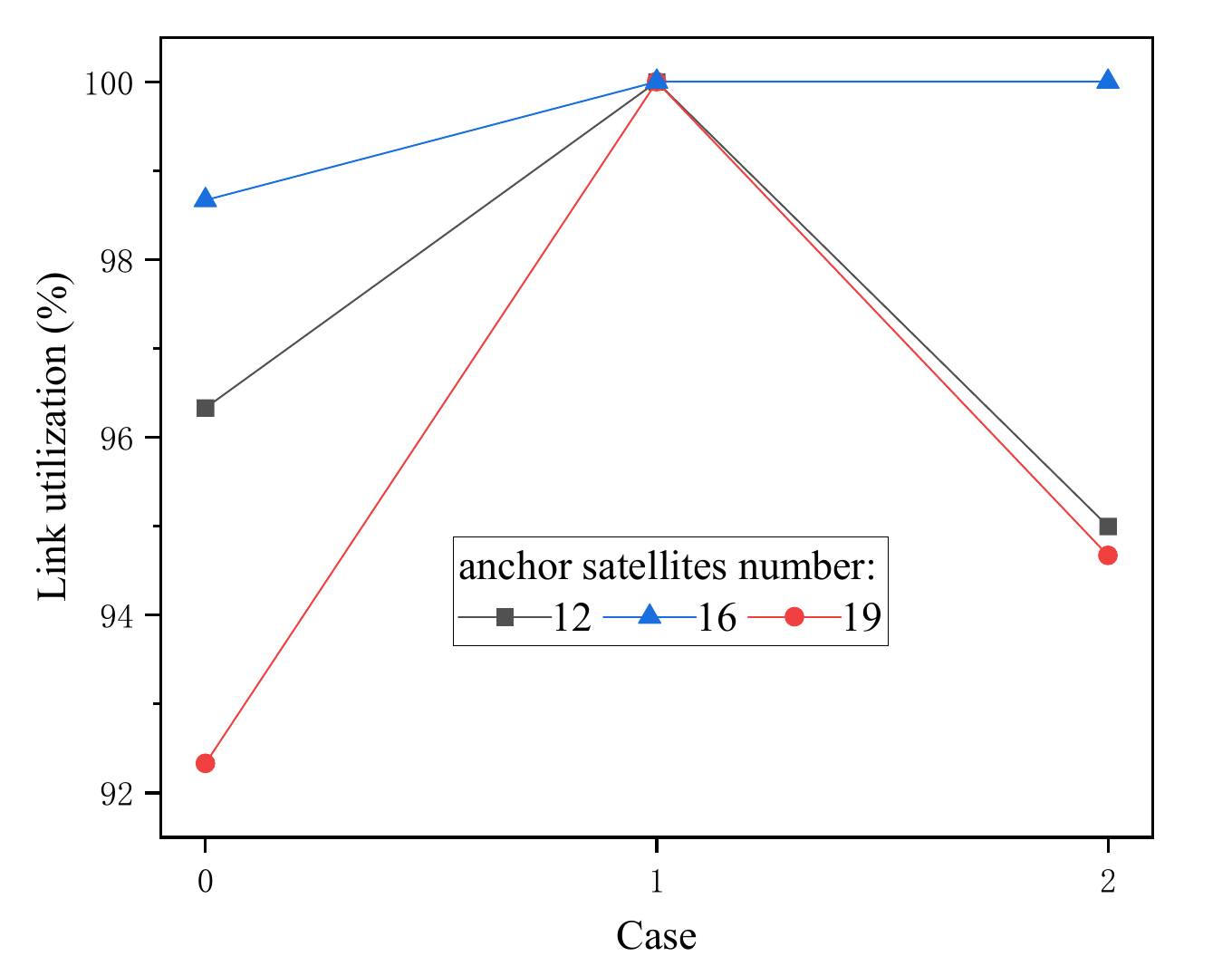}
		\caption{Link utilization in a single superframe.}
		\label{fig8}
	\end{figure}

\begin{figure}[] 
		\centering
		\includegraphics[width=0.8\linewidth]{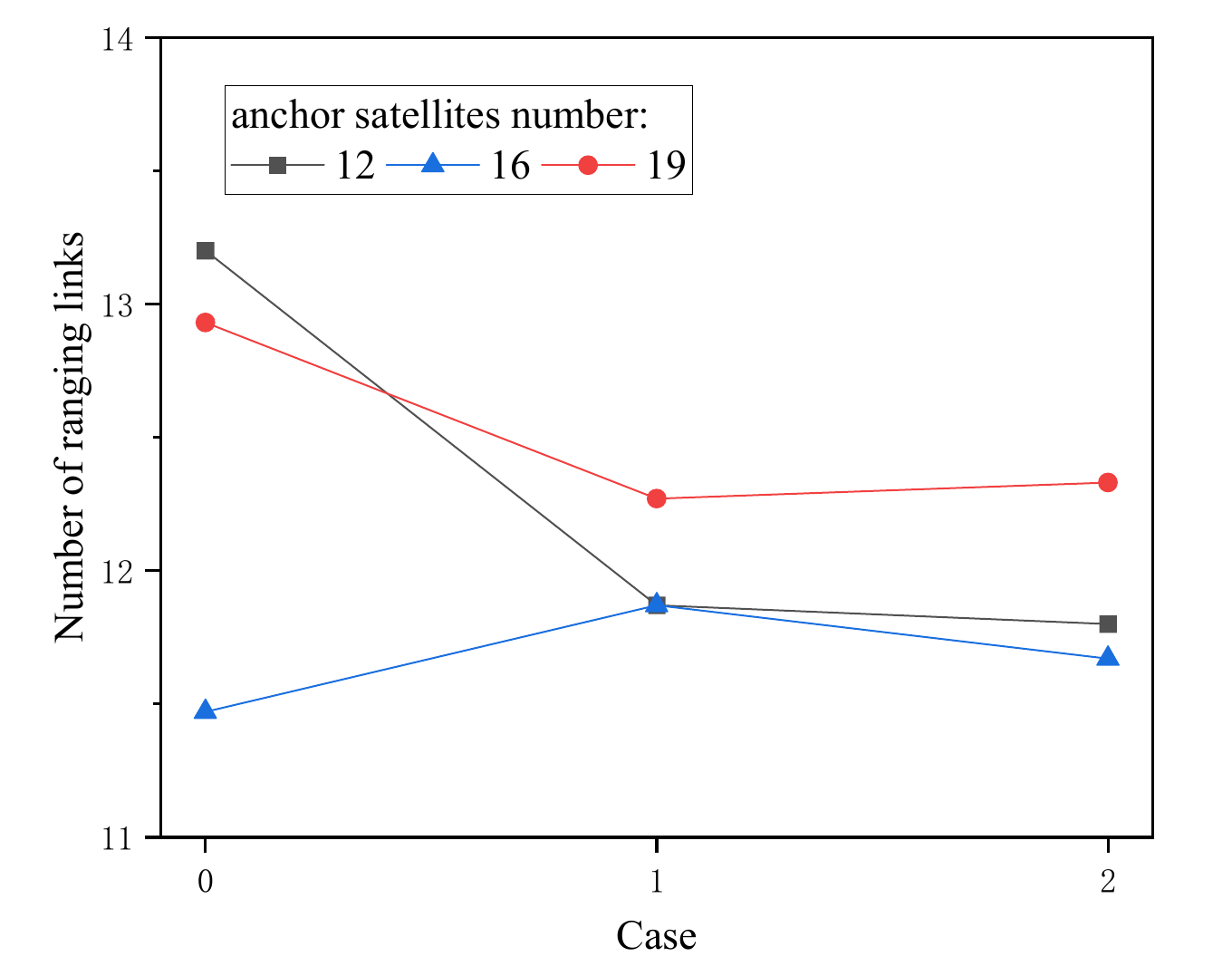}
		\caption{Number of ranging links in a single superframe.}
		\label{fig9}
	\end{figure}

	Fig.~\ref{fig6} to Fig.~\ref{fig9} illustrate the performance comparison of the GNSS within a superframe under different $\overrightarrow{{{U}_{s}}}$ configuration scenarios. On the x-axis:
	\begin{itemize}
		\item \textbf{Case 0}: Represents the scenario with no user demand, i.e., $\overrightarrow{{{U}_{s}}}=\{\}$.
		\item \textbf{Case 1}: Represents the scenario reaching the expansion capacity as shown in the first test case in Table~\ref{GNSS service capability}, i.e., $\overrightarrow{{{U}_{s}}}=\overrightarrow{{{U}_{s1}}}$.
		\item \textbf{Case 2}: Represents the scenario reaching the expansion capacity as shown in the second test case in Table~\ref{GNSS service capability}, i.e., $\overrightarrow{{{U}_{s}}}=\overrightarrow{{{U}_{s2}}}$.
	\end{itemize}
	
	\paragraph{Throughput}
	Figure~\ref{fig6} illustrates that, compared to scenarios without external users, the throughput of GNSS, when operating at its expansion capacity, generally exhibits either a stable or declining trend. 
	A stable trend suggests that incorporating throughput into the ILP optimization objective enables the GNSS to serve external users while minimizing the encroachment on time slot resources allocated between non-anchor and anchor satellites, thereby achieving the highest optimization objective value. 
	In contrast, a declining trend arises when the penalty for unmet user demands in the ILP optimization objective becomes significant, forcing the GNSS to encroach upon the time slots originally designated for non-anchor and anchor satellites to fulfill user service requirements.
	
	Additionally, The throughput varies across states with different numbers of anchor satellites.
	The highest throughput is achieved with 16 anchor and 14 non-anchor satellites. This configuration yields the maximum throughput because the nearly equal numbers ensure sufficient link opportunities for all non-anchor satellites.
	Throughput decreases with 12 anchor and 18 non-anchor satellites. Here, the excess of non-anchor satellites over anchors reduces the likelihood of each non-anchor satellite securing a link with anchor. The lowest throughput occurs with 19 anchor and only 11 non-anchor satellites. Although link opportunities are abundant per non-anchor satellite, the total potential link count is fundamentally limited by the small number of non-anchor satellites.

	\paragraph{Delay}
	Figure~\ref{fig7} illustrates the delay experienced by non-anchor satellites within a superframe. Compared to scenarios without external users, the delay generally increases as GNSS reaches its expansion capacity. When the number of anchor satellites is 12, the throughput in case 0 and case 1, as shown in Figure~\ref{fig6}, is identical. However, in Figure~\ref{fig7}, the delay in case 1 exhibits a slight decrease compared to case 0. Conversely, when the number of anchor satellites is 16, the throughput in case 1 and case 2 from Figure~\ref{fig6} remains the same, yet the delay in case 2 is higher than that in case 1 in Figure~\ref{fig7}.
	
	This observation highlights that while throughput and non-anchor satellite delay are related, they are not equivalent optimization objectives. 
	Higher throughput implies more frequent link establishments between non-anchor and anchor satellites, which should result in lower delays for non-anchor satellites. However, when differences in throughput are minimal, variations in the distribution of link establishments can lead to slight differences in delay.
	
	\paragraph{Link Utilization}
	The link utilization of a single satellite is defined as the ratio of the number of time slots during which its ISLs are active to the total number of time slots in a given time period. The constellation-wide link utilization is then derived by averaging the utilization of all satellites.
	Figure~\ref{fig8} illustrates the link utilization of the GNSS constellation. 
	In case 1, the link utilization reaches 100\%, while in case 2, it is lower compared to case 0. 
	This reduction in case 2 occurs because GNSS must provide users with ISLs lasting for two time slots. 
	These longer-duration ISLs inherently reduce the efficiency of GNSS resource scheduling, as they limit the flexibility to allocate link resources across the constellation.
	
	\paragraph{Ranging Links}
	Figure~\ref{fig9} depicts the number of ranging links among satellites in the GNSS constellation. The number of ranging links, which serves as a constraint condition, exhibits a fluctuating trend across different cases but consistently remains above the constraint of $L_{min} = 11$.
	
	Notably, compared to Case 0, which represents scenarios with no external users, Cases 1 and 2—achieving extended capacity—maintain relatively low latency and ensure a sufficient number of ranging links within the GNSS constellation. This outcome is attributed to the ILP-based GNSS CPD, which enforces the necessary communication and ranging resources for GNSS as hard constraints, thereby guaranteeing their availability despite external user demands.
	
	\subsubsection{Complete Cycle}
	
	This section analyzes the delay and ranging performance of the GNSS constellation over seven days (2016 FSA states) under several conventional cases. 
	Since no existing algorithm currently supports the provision of continuous multi-time-slot ISLs to external users, we compare the proposed ILP-based GNSS CPD with the FCP~\cite{38} approach in two scenarios: one where no ISLs are required for users and another where only single time-slot ISLs are provided.
	To adapt FCP to this context, we modify it such that once all satellites have provided the required number of ISLs for a specific user, FCP ceases allocating additional user-directed ISLs. 
	For the scenario requiring continuous multi-time-slot ISLs for users, a direct comparison of CPD methods is not performed. Instead, we conduct an intra-case comparative analysis to emphasize performance variations under these specific conditions.
	
	We have established three conventional cases, denoted as case1*, case2*, and case3*, with the user demands for each case detailed in Table~\ref{case description}. 
	In case1*, there are no external users. 
	In case2*, four GEO and IGSO users are included. 
	Case3* builds upon case2* by adding one L3, one L4, one L5, and one lunar DRO user.
	
	\begin{table}[]
		\caption{Case Description}
		\label{case description}
		\centering
		\footnotesize
		\renewcommand{\arraystretch}{1.2}
		\begin{tabular}{|c|c|}
			\hline
			\thead{Case} & \thead{Description of $\vec{U}$ in Each Case} \\
			\hline
			case1* &  $\vec{U}$ =\{\}  \\
			\hline
			case2* & \thead{$\vec{U}$=\{($U_{GEO_1}$,~[1,1,4,1]), \\($U_{GEO_2}$,~[1,1,4,1]),\\($U_{IGSO_1}$,~[1,1,4,1]),\\ ($U_{IGSO_2}$,~[1,1,4,1])\}}\\
			\hline
			case3* & $\vec{U}$=$\vec{U}$ in case2* + \thead{~\{($U_{L3_1}$,~~[1,2,4,1]), \\($U_{L4_1}$,~[1,2,4,1]),\\($U_{L5_1}$,~[1,2,4,1]),\\ ($U_{DRO_1}$,~[1,2,4,1])\}} \\
			\hline
		\end{tabular}
	\end{table}
	
\begin{figure}[] 
		\centering
		\includegraphics[width=0.8\linewidth]{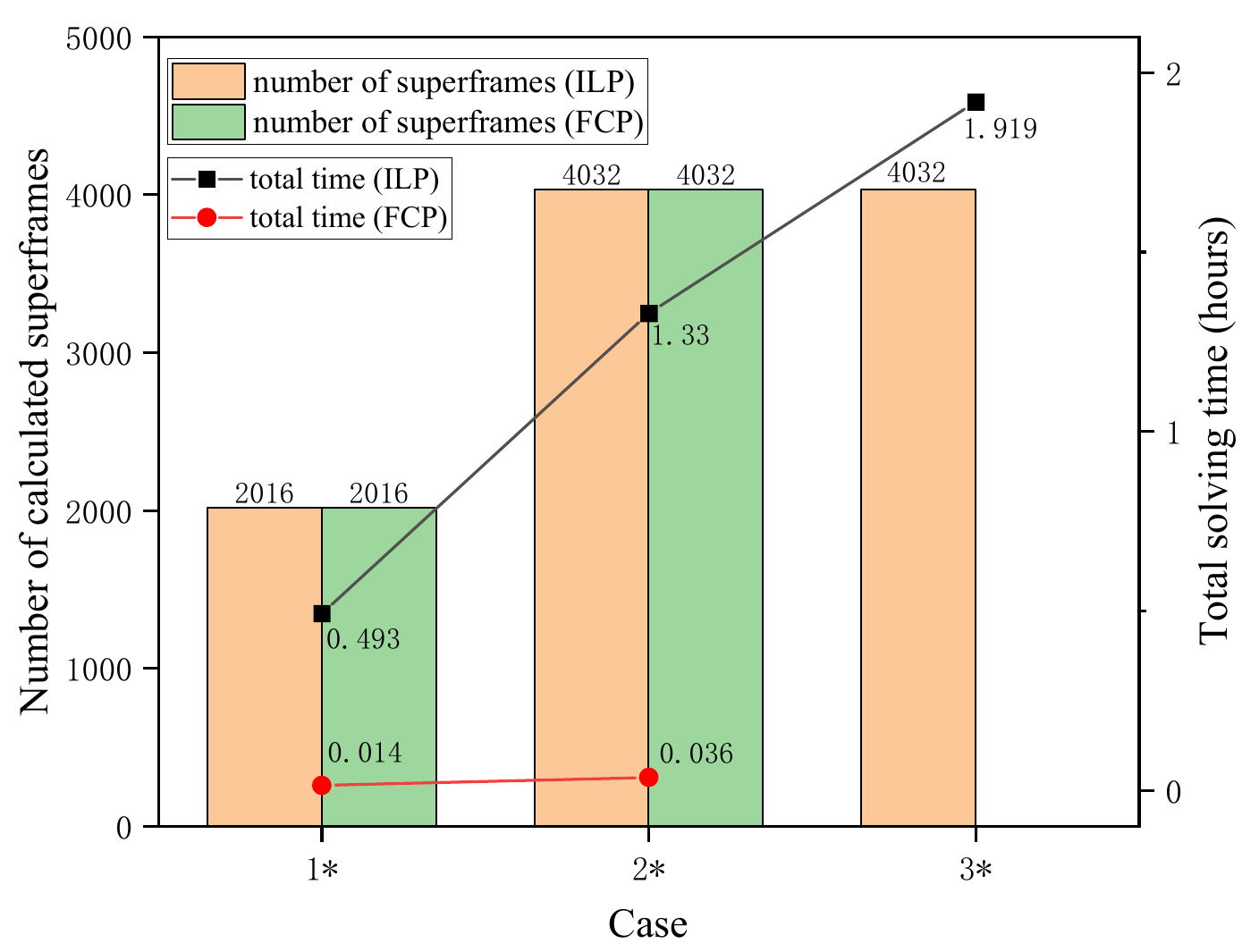}
		\caption{Comparison of solving time and number of calculated superframes.}
		\label{fig10}
	\end{figure}

	\paragraph{Solving Time}
	Fig.~\ref{fig10} illustrates the total solving time and the number of superframes calculated for ILP-based GNSS CPD and FCP across three cases.
	Since FCP lacks the capability to provide continuous multi-time-slot ISLs, it cannot operate under Case 3*, leading to missing data for FCP in this scenario.
	Overall, the solving time for FCP is significantly shorter than that for ILP; however, the solution time for ILP remains acceptable relative to the total duration of the link scheduling period (i.e., seven days). 
	The solving time for Case 1* is the shortest, with only 2016 superframes computed, indicating that a single superframe's link scheduling is calculated per FSA state. Cases 2* and 3* require 4032 superframes, signifying that two superframes' link scheduling is calculated for all FSAs. 
	
	In Cases 2* and 3*, the first superframe satisfies all user requirements, while the second plans only the internal ISLs of GNSS, thereby ensuring a 100\% user satisfaction ratio in both cases. The solving time for ILP increases from Case 2* to Case 3*, a trend attributed to the fact that as more constraints are added to the ILP problem, the difficulty of solving it generally increases, thus requiring additional time to find a solution.
	
\begin{figure}[] 
		\centering
		\includegraphics[width=0.8\linewidth]{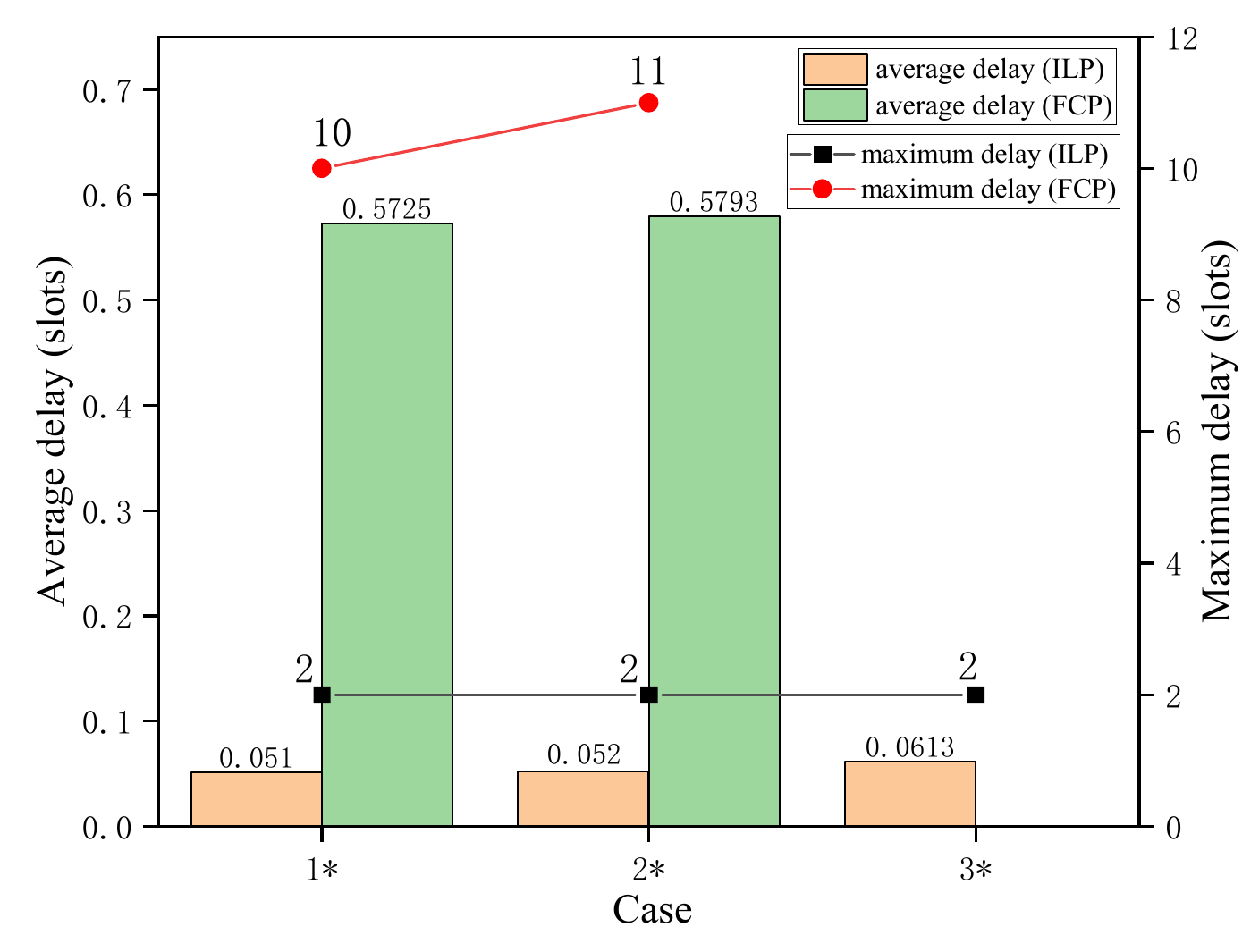}
		\caption{Comparison of delay.}
		\label{fig11}
	\end{figure}

	\paragraph{Delay}
	Fig.~\ref{fig11} illustrates the average and maximum delay for non-anchor satellites across three cases. 
	Both FCP and ILP-based GNSS CPD achieve an average delay of less than 1 slot, with ILP's average delay approaching nearly 0 slots. 
	This performance is primarily attributed to the fact that, in most FSA states, the number of anchor satellites exceeds that of non-anchor satellites. 
	Consequently, non-anchor satellites have ample opportunities to establish ISLs with anchor satellites during the planning process, resulting in low delay values.
	In the ILP model, constraints are imposed on the maximum delay for non-anchor satellites, ensuring it does not exceed 2 slots. In contrast, FCP focuses on fairness in link establishment between nodes, without distinguishing between anchor and non-anchor satellites. 
	This leads to a maximum delay of up to 11 slots for non-anchor satellites.
	
	Overall, ILP demonstrates significantly superior performance in terms of delay compared to FCP.
	
\begin{figure}[] 
		\centering
		\includegraphics[width=0.8\linewidth]{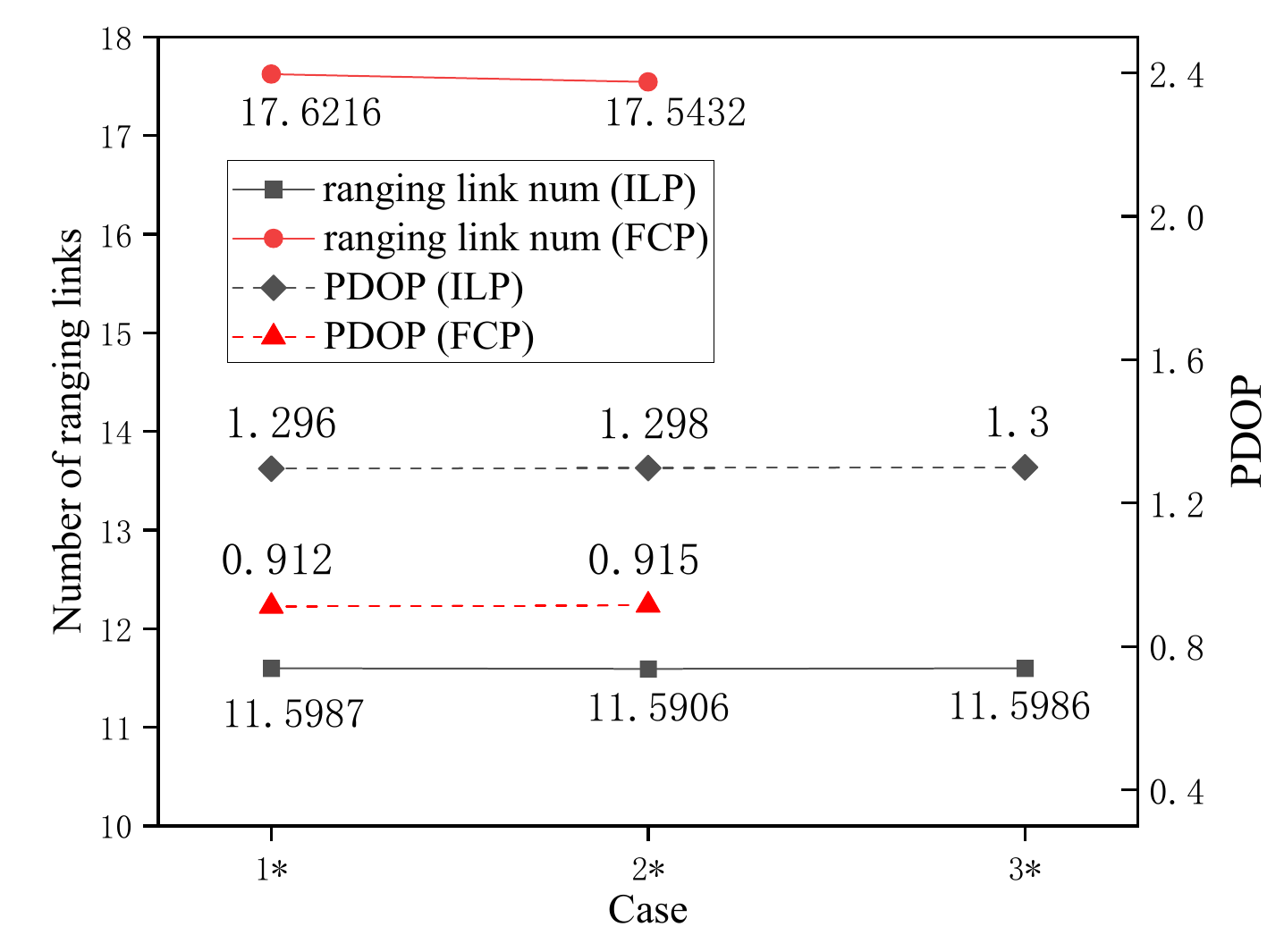}
		\caption{Comparison of number of ranging links and PDOP value.}
		\label{fig12}
	\end{figure}

	\paragraph{Ranging Links}
	Fig.~\ref{fig12} compares the average number of ranging links and PDOP between FCP and ILP-based GNSS CPD. While FCP demonstrates a higher average number of ranging links compared to ILP, the ILP consistently maintains the average number of ranging links above the minimum constraint of $L_{min}=11$. 
	
	Overall, FCP exhibits superior performance regarding the total number of ranging links and PDOP. 
	However, the ILP-based method achieves a superior PDOP value below 3. This level of precision is sufficient to ensure the high orbit determination performance of the GNSS~\cite{27}.

\begin{figure}[] 
		\centering
		\includegraphics[width=0.8\linewidth]{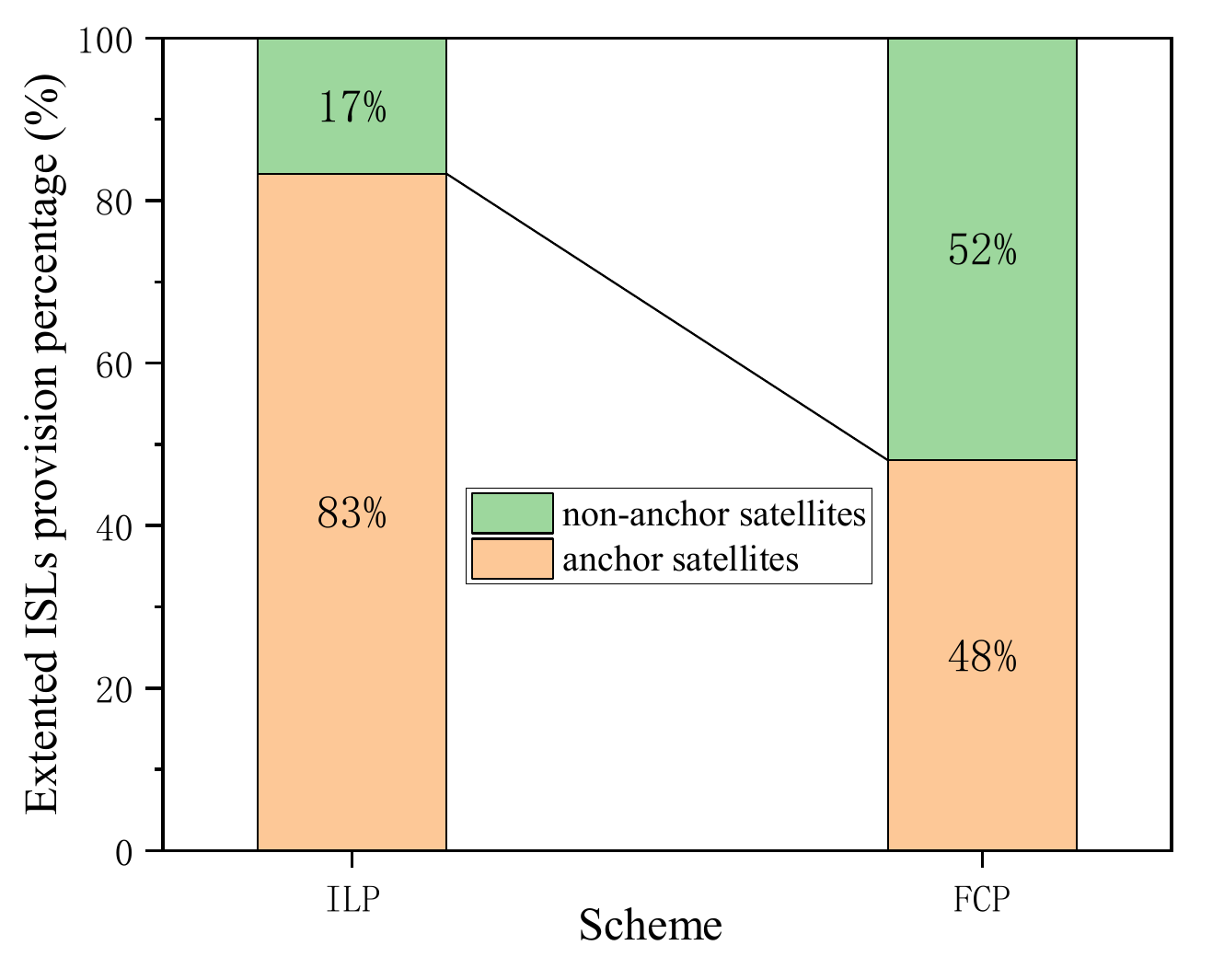}
		\caption{Comparison of extended ISLs provision percentage.}
		\label{fig13}
	\end{figure}

	\paragraph{Extended ISLs provision pecentage }
	Fig.~\ref{fig13} illustrates the proportion of ISLs provided to users by anchor and non-anchor satellites under Case 2* for both the ILP-based GNSS CPD and FCP methods. 
	For FCP, which inherently does not distinguish between anchor and non-anchor satellites, the higher proportion of ISLs provided by non-anchor satellites is likely due to richer visibility opportunities between users and non-anchor satellites or the randomness inherent in its underlying matching algorithm. 
	In contrast, the ILP-based approach prioritizes anchor satellites for providing ISLs when both anchor and non-anchor satellites are capable of doing so. 
	This prioritization stems from the need to satisfy constraints on the maximum delay for non-anchor satellites and to maximize throughput as part of the optimization objective. 
	As a result, the proportion of ISLs provided by anchor satellites in the ILP-based method is significantly higher than that provided by non-anchor satellites.
	
	\subsection{BeiDou ISL Impact on the Earth-Moon Libration Point Constellation}
	\label{eval_part3}
	Compared to GSs, utilizing GNSS ISL for OD of LP constellations offers the following advantages:
	\begin{enumerate}
		\item In the event of GS failures, GNSS can serve as a backup facility for navigating (or timing) LP constellations.
		\item Due to siting constraints and Earth’s rotation, GSs may not maintain continuous visibility with LP users, thus failing to provide continuous navigation and timing services. In contrast, GNSS systems such as BeiDou, with their widely distributed satellite orbits and large number of satellites, ensure that at least some GNSS satellites remain visible to LP constellations at all times.
		\item  GNSS satellites operate in medium to high Earth orbits (above 20000 km from the geocenter). When conducting ranging with LP users, the ranging geometry of GNSS satellites is superior to that of GSs (located at 6371 km from the geocenter), which contributes to faster convergence in OD.
		\item Unlike GSs, the ranging links between GNSS satellites and LP users do not pass through Earth's atmosphere, thereby improving ranging accuracy.
	\end{enumerate}
	These advantages motivate the experiments conducted in this section.
	Specifically, this section examines the influence of GNSS ISLs on the clock bias and OD of the Earth-Moon LP constellation deployed at the L3, L4, L5 points, and in lunar DRO~\cite{19}. 
	During the GNSS CPD process, users located at these LPs and in DRO are treated as a single logical user entity, $U_l$ (corresponding to 'User Preprocess' in Sec~\ref{sec_system_model}-B-1).
	
	After obtaining the GNSS CPD results, the LP constellation evaluates whether to accept or reject the ISLs offered by GNSS. 
	In this simulation, the strategy is to accept all ISLs provided by GNSS, with the specific receiving node being the LP satellite that: 1): Has visibility to the GNSS satellite offering the ISL, and 2): Possesses the minimum satellite ID among the visible satellites (corresponding to 'Isolation between GNSS and User' in Sec~\ref{sec_system_model}-B-2). 
    For example, we assign the satellite IDs for L3, L4, L5, and DRO satellite as 1, 2, 3, and 4, respectively. If both L4 and L5 are visible to the GNSS, then the L4 satellite is selected as the receiving satellite for the GNSS ISL.
	
	Once the receiving node within the LP constellation is selected, its visibility to all other entities is set to zero during the intervals when the GNSS provides ISLs. 
	Based on this adjusted visibility, the LP constellation performs its internal CPD to ensure that each LP satellite establishes a link with all other visible LP satellites at least once every five minutes. 
	This methodology ensures the compatibility of the CPD process between GNSS and the LP constellation.
	
	The OD method employed in this paper is a batch processing approach combined with least-squares optimization of the orbit state vectors and clock bias parameters~\cite{56}. 
	The coordinate system used is the Earth-Centered Inertial frame. 
	In navigation systems, an accurate antenna phase center offset model is typically established in advance. This model provides correction values that account for variations of the antenna phase center. 
	By applying this model during post-processing in OD algorithms, nearly all inherent phase errors can be eliminated; therefore, ranging errors caused by antenna phase center variations are not considered in this experiment. 
	Furthermore, since both GNSS satellites and libration point satellites operate well above Earth's atmosphere, ionospheric and tropospheric delays do not contribute to range errors. 
	The primary error sources considered in this experiment are random ranging errors induced by Gaussian white noise and systematic errors resulting from minor environmental-induced variations in internal instrumentation, such as antenna feed lines.
	Specifically, the ranging random error is set to 2 m, with a systematic error of 1 m. 
	The initial orbit error is 10,000 m, and the initial velocity error is 1 m/s. The arc length for OD is set to 3 days. 
	The OD results are summarized in Tables~\ref{clock error},~\ref{position error}, and~\ref{velocity error}.
	This paper assumes the GNSS clocks are accurate; therefore, the clock error in the LP constellation is relative to the GNSS clocks. 	
	In the tables:
	
	\begin{table}[t]
		\caption{Clock error}
		\label{clock error}
		\centering
		\scalebox{0.9}{ 
		\begin{tabular}{|c|c|c|c|}
			\hline
			& \thead{$\vec{U}$=\{\}}&\thead{$\vec{U}$ =\\ \{$U_l$, [1,2,1,1]\}}&\thead{$\vec{U}$=\\ \{$U_l$, [1,2,3,1]\}}\\
			\hline
			L3 Clock error&1.7849 ns & 0.3076 ns& 0.1378 ns \\
			\hline
			L4 Clock error&1.9019 ns &0.3974 ns&0.1187 ns\\
			\hline
			L5 Clock error&1.7304 ns & 0.1609 ns &0.2252 ns\\
			\hline
			DRO Clock error	&2.0101 ns &0.2101 ns&0.2556 ns\\
			\hline
			
		\end{tabular}
	}
	\end{table}
	
	\begin{table}[t]
		\caption{Position error}
		\label{position error}
		\centering
		\scalebox{0.9}{ 
		\resizebox{0.49\textwidth}{!}{
		\begin{tabular}{|c|c|c|c|}
			\hline
			& \thead{$\vec{U}$=\{\}}&\thead{$\vec{U}$ =\\ \{$U_l$, [1,2,1,1]\}}&\thead{$\vec{U}$=\\ \{$U_l$, [1,2,3,1]\}}\\
			\hline
			L3 X Pos. error&219.439 m & 1.605 m& 0.555 m \\
			L3 Y Pos. error&959.650 m & 0.589 m&1.143 m \\
			L3 Z Pos. error&515.600 m  & 1.756 m& 1.032 m \\
			L3 Pos. error &1111.271m  & 2.451 m & 1.637 m\\
			\hline
			L4 X Pos. error&26.633 m& 4.786 m& 3.263 m \\
			L4 Y Pos. error&616.564 m& 12.380 m& 9.107 m \\
			L4 Z Pos. error&596.172 m & 11.753 m& 8.093 m \\
			L4 Pos. error & 858.068 m &17.729 m & 12.613 m \\
			\hline
			L5 X Pos. error&105.147 m & 1.204 m& 1.025 m \\
			L5 Y Pos. error&810.522 m & 1.253 m& 2.424 m \\
			L5 Z Pos. error&613.651 m & 2.117 m& 3.625 m \\
			L5 Pos. error &1022.042 m &2.739 m & 4.479 m  \\
			\hline
			DRO X Pos. error&16.741 m & 0.536 m& 0.614 m \\
			DRO Y Pos. error&3.623 m &1.553 m  & 0.454 m \\
			DRO Z Pos. error&74.468 m & 1.130 m& 0.445 m \\
			DRO Pos. error &  76.413 m & 1.994 m & 0.883 m     \\
			\hline
			
		\end{tabular}}
	}
	\end{table}
	
	\begin{table}[t]
		\caption{Velocity error}
		\label{velocity error}
		\centering
		\scalebox{0.9}{ 
		\resizebox{0.49\textwidth}{!}{
		\begin{tabular}{|c|c|c|c|}
			\hline
			& \thead{$\vec{U}$=\{\}}&\thead{$\vec{U}$ =\\ \{$U_l$, [1,2,1,1]\}}&\thead{$\vec{U}$=\\ \{$U_l$, [1,2,3,1]\}}\\
			\hline
			L3 X Vel. error&1.254 mm/s & 6 um/s& 2 um/s \\
			L3 Y Vel. error&0.839 mm/s & 3 um/s &7 um/s \\
			L3 Z Vel. error&3.034 mm/s  & 16 um/s& 3 um/s \\
			L3 Vel. error&3.632 mm/s  & 17.349 um/s& 7.874 um/s \\
			\hline
			L4 X Vel. error&0.211 mm/s& 15 um/s& 13 um/s \\
			L4 Y Vel. error&0.991 mm/s& 13 um/s& 6 um/s \\
			L4 Z Vel. error&0.490 mm/s& 10 um/s& 4 um/s \\
			L4 Vel. error&1.125 mm/s  & 22.226 um/s& 14.866 um/s \\
			\hline
			L5 X Vel. error&0.635 mm/s & 14 um/s& 8 um/s \\
			L5 Y Vel. error&0.453 mm/s & 17 um/s& 10 um/s \\
			L5 Z Vel. error&0.573 mm/s & 9 um/s& 8 um/s \\
			L5 Vel. error&0.968 mm/s  & 23.791 um/s& 15.099 um/s \\
			\hline
			DRO X Vel. error&0.201 mm/s & 12 um/s& 9 um/s \\
			DRO Y Vel. error&0.056 mm/s & 23 um/s& 7 um/s \\
			DRO Z Vel. error&1.047 mm/s & 7 um/s& 2 um/s \\
			DRO Vel. error&1.068 mm/s  & 26.870 um/s& 11.576 um/s \\
			\hline
			
		\end{tabular}}
	}
	\end{table}
	\begin{itemize}
		\item $\vec{U} = \{\}$ indicates that GNSS does not provide any link establishment to the LP logical user $U_l$.
		\item $\vec{U} = \{U_l, [1,2,1,1]\}$ indicates that within each FSA, GNSS provides one instance of ISL with a duration of two time slots to the LP user $U_l$ equipped with a single antenna terminal.
		\item $\vec{U} = \{U_l, [1,2,3,1]\}$ indicates that within each FSA, GNSS provides three instances of ISL with a duration of two time slots to the LP user $U_l$ equipped with a single antenna terminal.
	\end{itemize}
	
	This setup enables an analysis of how varying frequencies of ISL support from GNSS affect the performance and accuracy of OD for the LP constellation.
	In the case denoted by $\vec{U} = \{\}$, where no GNSS ISL support is provided, calculating precise absolute clock bias values for the LP constellation becomes challenging, requiring a priori clock bias estimates. 
	As a result, both clock bias accuracy and OD accuracy are significantly degraded.
	With GNSS ISL support, clock bias and OD accuracy improve substantially compared to scenarios without such support. 
	In general, higher frequencies of GNSS-provided ISLs yield more precise results. 
	However, due to system biases and random errors, there are instances where higher-frequency ISL support may produce slightly less accurate results than lower-frequency ISL support. This phenomenon is attributed to the effects of ranging errors and is considered normal.
	
	These findings highlight the critical role of GNSS ISL support in enhancing the precision of clock bias estimation and OD for the LP constellation.
	
	\subsection{Discussion}
	Based on the aforementioned simulation results, we draw the following conclusions:
	
	\begin{enumerate}
		\item The proposed ILP-based GNSS CPD maintains stable delay and ranging performance for the GNSS system, even under extended capacity conditions. This demonstrates the robustness and reliability of the ILP-based approach.
		\item While FCP exhibits adaptability in scenarios requiring only single-slot extended ISLs, it cannot provide multi-slot ISLs to external users. 
		Additionally, it may result in high maximum delays for non-anchor satellites. 
		In contrast, the ILP-based GNSS CPD effectively accommodates multi-slot extended ISLs and enforces strict constraints on the maximum delay for non-anchor satellites. This capability enhances real-time interaction between GSs and non-anchor satellites.
		\item In scenarios involving single-slot extended ISLs, the ILP-based GNSS CPD demonstrates a more rational service structure than FCP. 
		The ILP-based approach prioritizes anchor satellites for providing ISLs to external users, whereas FCP does not differentiate between anchor and non-anchor satellites, leading to suboptimal resource allocation.
		\item The Earth-Moon LP constellation benefits significantly from GNSS ISL support, with marked improvements in ranging accuracy and clock synchronization precision. 
		This underscores GNSS's potential as a central PNT infrastructure within a comprehensive PNT framework, supporting other PNT infrastructures in establishing and maintaining a precise spatiotemporal reference.
	\end{enumerate}
	
	\section{Conclusions}
	\label{sec_conclusion}
	This paper presents the CPD scheme as a robust framework for extending the application capabilities of GNSS ISLs. 
	We first analyze the feasibility of leveraging GNSS ISLs for broader use cases and propose a service process tailored to meet diverse external extension requirements within the GNSS framework.
	The GNSS CPD is formulated as an ILP model incorporating critical ranging and communication resource constraints, aiming to maximize system throughput while addressing expanded operational needs.
	
	A key innovation of this model is its ability to deliver continuous multi-time-slot ISLs within a polling time-division structure, effectively ensuring service continuity in a time-divided framework. 
	Simulation results confirm the model’s strong performance under both extended capacity and routine operational scenarios. 
	Moreover, the study highlights the significant impact of GNSS ISLs on the Earth-Moon LP constellation, demonstrating enhancements in orbit determination accuracy and clock synchronization precision.
	
	The proposed CPD scheme paves the way for expanding GNSS applications, establishing GNSS as a more pervasive, integrated, and interoperable PNT infrastructure capable of supporting future space and terrestrial needs.
	
	\bibliographystyle{IEEEtran}
	\bibliography{references}

\end{document}